\documentclass[12pt]{article}
\usepackage{epsf}

\newcommand{\be}{\begin{equation}}
\newcommand{\ee}{\end{equation}}

\newcommand{\tr}{{\rm Tr}}

\newcommand{\rmd}{{\rm d}}
\newcommand{\veck}{\vec k}
\newcommand{\vecp}{\vec p}

\begin{document}

\title{Derivative Expansion and the Effective Action for the Abelian
Chern-Simons Theory at Higher Orders}

\author{F. T. Brandt$^a$, Ashok Das$^b$, J. Frenkel$^a$ and J. C. Taylor$^c$
\\ \\
$^a$Instituto de F\'{\i}sica,
Universidade de S\~ao Paulo\\
S\~ao Paulo, SP 05315-970, BRAZIL\\
$^b$Department of Physics and Astronomy\\
University of Rochester\\
Rochester, NY 14627-0171, USA\\
$^c$ Department of Applied Mathematics\\ 
and Theoretical Physics\\
University of Cambridge\\
Cambridge, UK}
\date{}
\maketitle

\begin{abstract}

We study systematically the higher order corrections to the parity
violating part of the effective action for the Abelian Chern-Simons
theory  in $2+1$ dimensions, using the method of derivative
expansion. We explicitly calculate the parity violating parts of the
quadratic, cubic and the quartic terms (in fields) of the effective
action. We show that each of these actions can be summed, in
principle, to all orders in the derivatives. However, such a structure
is complicated and not very useful. On the other hand, at every order
in the powers of the derivatives, we show that the effective action
can also be summed to all orders in the fields. The resulting actions
can be expressed in terms of the leading order effective action in the
static limit. We prove gauge invariance, both {\it large} and {\it
small} of the resulting effective actions. Various other features of
the theory are also brought out.

\end{abstract}
\vfill\eject

\section{Introduction:}

Chern-Simons (CS) theories, in $2+1$ dimensions, have been of interest in
recent years \cite{1,2}. Non-Abelian CS theories are invariant under {\it large
gauge} transformations provided the coefficient of the CS term is
quantized. At finite temperature, however, the induced CS term has a
continuous coefficient (temperature dependent), which is incompatible
with the discreteness of the CS coefficient necessary for {\it large
gauge} invariance to hold \cite{3}. This puzzle of violation of {\it large
gauge} invariance at finite temperature is well understood now, at
least in the Abelian theory \cite{4,5,6,7}.

As is well known, at finite temperature, amplitudes \cite{8'} as well as the
effective action become non-analytic \cite{8,9}, unlike at zero
temperature.  As a
result, it  becomes essential to talk of the effective action only in
certain limits - the conventional ones being the long wave and the
static limits. It has already been shown within the context of the
Abelian CS theory in $2+1$ dimensions that {\it large gauge} invariance
is not an issue in the long wave limit \cite{7}. On the other hand, truncation
of the effective action at any finite order in the static limit leads
to violation of {\it large gauge} invariance although the complete
action has {\it large gauge} invariance. This has been checked for the
leading order terms in the effective action in the static limit.

The leading order parity violating effective action in the static
limit can be determined exactly, either through functional methods
\cite{6}  or through
the use of the {\it large gauge} Ward identity \cite{7,10}. This, of course,
raises the question of the higher order corrections to this action and
the issue of {\it large gauge} invariance for such terms. In this
paper, we address this question.

The higher order corrections can be naturally obtained through a
derivative expansion (powers of momentum) \cite{3,11,12}. However, as is
known in
simple models, in a model with {\it large gauge} invariance,
derivative expansion does lead to new subtleties \cite{13,14}. In
fact, some  such
subtlety was already noted earlier in the Abelian $2+1$ dimensional
model \cite{13}, even at the leading order in the static limit.
In section {\bf 2}, we extend and
improve on the analysis of \cite{13} to rederive, within the context of
derivative expansion, the leading order parity violating effective
action in the static limit which is linear in $B$, the magnetic
field. We also show that the parity violating part of the effective
action does not contain any higher order terms in $\vec{A}$ so that
this action is the complete parity violating effective action in that
limit, consistent with the results obtained in \cite{5,6,7}. In
section  {\bf 3}, we tackle the question of going beyond the
leading order and evaluate the parity violating effective action from
a calculation of amplitudes in momentum space. The two point (self-energy)
as well as the four point (box diagram) amplitudes are evaluated
explicitly to fifth order in the power of momentum. However, momentum space
calculations become increasingly more involved as we go to higher
orders.  As a result, in
section {\bf 4}, we use the derivative expansion in the coordinate
space. Here, too, we calculate a closed form expression for the
quadratic part of the  parity violating effective action. In fact, the
effective action, at any order, can be obtained in a closed form, but
the closed form expressions are not necessarily simple. Rather, a
power series expansion in the number of derivatives gives a simpler
expression to the quadratic, cubic and quartic (in fields) terms of
the effective action. One of the features that arises from this
analysis is the fact that the expressions for these actions are not
manifestly invariant under {\it small} gauge transformations, although
they can be brought  to a gauge invariant form (at
least for lower orders) through the use of various algebraic
identities. In section {\bf 5}, therefore, we give a general proof of
gauge invariance of the effective action under {\it small} gauge
transformations and  recast the derivative
expansion into a gauge invariant form. In section {\bf 6}, we analyze
the general features of our results. In particular, we show that from
our low order (in fields) calculations, we can, in fact, predict the
behaviour of the effective action with one, three and five derivatives
to all orders in the fields. In fact, to all orders, we find that
these effective actions are completely determined from the form of the
leading order parity violating effective action in the static
limit. They are manifestly invariant under {\it small} as well as {\it
large} gauge transformations. We present a brief conclusion in section {\bf 7}. 

\section{Leading Order Derivative Expansion in the Static Limit:}

Let us consider a fermion interacting with an Abelian gauge background
described by the Lagrangian density (in $2+1$ dimensions)

\begin{equation}\label{lagra}
{\cal L} = \overline{\psi}\left(\gamma^{\mu}(i\partial_{\mu} +
eA_{\mu}) - M\right)\psi
\end{equation}
Here, for simplicity, we will assume that $M > 0$. The effective
action following from this is formally given by

\newpage

\begin{eqnarray}
\Gamma_{eff}[A, M] & = & -i \ln \det\left(\gamma^{\mu}(i\partial_{\mu} +
eA_{\mu}) - M\right)\nonumber\\
 & = & -i {\rm Tr}
\ln\left(\gamma^{\mu}(i\partial_{\mu} + eA_{\mu}) - M\right)
\end{eqnarray}
where \lq\lq Tr'' stands for trace over Dirac indices as well as over
a complete basis of states. As is well known, at finite temperature,  
the effective action is not well defined everywhere \cite{8,9}, as a
result of which it
can be expanded in powers of derivatives only  in some
limit. This is a simple reflection of the fact that thermal amplitudes are
non-analytic at the origin in the energy-momentum plane \cite{8'}. This was
explicitly shown for the $2+1$ dimensional Abelian Chern-Simons
theory \cite{7}, where we calculated the leading term in the parity
violating part of the box diagram at finite temperature and where we also
showed that {\it large gauge} invariance is an issue in the static
limit, but not in the long wave limit. In
this paper, we systematically calculate the higher order corrections
to the earlier result, in the static limit, by using the method of derivative
expansion \cite{11,12}. Although we had earlier summed the leading order terms in the
static limit using a large gauge Ward identity \cite{10}, in this section, we
will rederive this result from the derivative expansion.

The leading order behaviour in the static limit, as was shown in
\cite{7}, is consistent with assuming a specific form of the
background gauge fields, namely \cite{6}, 
\begin{equation}
A_{0} = A_{0}(t),\qquad \vec{A} = \vec{A}(\vec{x})\label{a1'}
\end{equation}
The parity violating part of the effective action, in such a background,
was already calculated in \cite{6,7,13} and here, as a warm up, we rederive the
result from a derivative expansion. It is well known that, by a
suitable gauge transformation \cite{6,7,15}, 
\begin{equation}
A_{\mu} \rightarrow A_{\mu} + \partial_{\mu}\Omega,\quad \Omega (t) =
\left(- \int_{0}^{t} +
{t\over\beta}\int_{0}^{\beta}\right)dt'\,A_{0}(t')\label{gauge}
\end{equation}
a static gauge background
of the  form (\ref{a1'}) can always be brought to the form 
\begin{equation}\label{adef}
A_{0}(t) \rightarrow \frac{a}{\beta} = \frac 1 \beta 
\int_0^\beta dt A_{0}(t),\qquad \vec{A} = \vec{A}(\vec{x})
\end{equation}
so that under a {\it large gauge} transformation,
\begin{equation}
a \rightarrow a  + {2\pi n\over e}\label{a1}
\end{equation}
We will use the imaginary time formalism \cite{8,16,17} in evaluating the finite
temperature determinant, where energy takes discrete values. Rotating
to Euclidean space, the effective action takes the form
\begin{equation}\label{a2}
\Gamma_{eff}[A,M] = - \sum_{n} {\rm Tr}\;\ln \left(p\!\!\!\slash +
\gamma_{0}\tilde{\omega}_{n} + M + eA\!\!\!\slash\right)
\end{equation}
where we have defined $p\!\!\!\slash = \vec{\gamma}\cdot \vec{p}$ and
similarly $A\!\!\!\slash = \vec{\gamma}\cdot \vec{A}$. Furthermore,
\begin{equation}
\tilde{\omega}_{n} = \omega_{n} + \frac{ea}{\beta} = {(2n+1)\pi\over \beta} +
\frac{ea}{\beta}\label{a3} 
\end{equation}
where $\beta$ represents the inverse temperature in units where the
Boltzmann constant is unity. The momentum in
the above expression is to be understood as an operator which does not
commute with coordinate dependent quantities. Let us also note that we
are working with the following representation of the gamma matrices in
the Euclidean space
\begin{equation}
\gamma_{0} = i\sigma_{3},\quad \gamma_{1} = i\sigma_{1},\quad
\gamma_{2} = i\sigma_{2}\label{a3''}
\end{equation}

Let us next define
\begin{equation}
K_{n} = {1\over p\!\!\!\slash + \gamma_{0}\tilde{\omega}_{n} + M}
\end{equation}
Then, taking out a normalization factor, we can write the effective
action as
\begin{equation}
\Gamma_{eff} [A,M] = - \sum_{n} {\rm Tr}\;\ln \left(1 + eK_{n}
A\!\!\!\slash\right) = - \sum_{n} {\rm Tr}\;\sum_{j=0}^{\infty}
{(-1)^{j}\over j+1} (e K_{n}A\!\!\!\slash)^{j+1}\label{a3'}
\end{equation}
This expression shows that the effective action contains all powers of
$\vec{A}$. However, let us next show that quadratic and higher powers
of $\vec{A}$ do not occur in the parity violating part of the
effective action. To that end, let us note that if we define \cite{6}
\begin{equation}
\gamma_{0}\tilde{\omega}_{n} + M =
\rho_{n}\,e^{\gamma_{0}\phi_{n}}\label{a4} 
\end{equation}
where
\begin{equation}
\rho_{n} = \sqrt{\tilde{\omega}_{n}^{2} + M^{2}},\qquad \phi_{n} =
\tan^{-1} {\tilde{\omega}_{n}\over M}\label{a5}
\end{equation}
we can write
\begin{eqnarray}
K_{n} & = & {1\over p\!\!\!\slash + \gamma_{0}\tilde{\omega}_{n} +
M}\nonumber\\
 & = & e^{-\gamma_{0}\phi_{n}/2}\; {1\over p\!\!\!\slash +
\rho_{n}}\;e^{-\gamma_{0}\phi_{n}/2} =
e^{-\gamma_{0}\phi_{n}/2}\;K_{n}^{(0)}\;e^{-\gamma_{0}\phi_{n}/2}
\end{eqnarray}

Using this, the terms in (\ref{a3'}) containing higher powers of $\vec{A}$ can
be written as
\begin{eqnarray}
\Gamma_{eff}^{(higher)} [A,M] & = & - \sum_{n} {\rm Tr}\,\sum_{j=1}
{(-1)^{j}\over j+1} (eK_{n}A\!\!\!\slash)^{j+1}\nonumber\\
 & = & - \sum_{n} {\rm Tr}\,\sum_{j=1}
{(-1)^{j}\over j+1}
(e^{-\gamma_{0}\phi_{n}/2}eK_{n}^{(0)}e^{-\gamma_{0}\phi_{n}/2}A\!\!\!\slash)^{j+1}\nonumber\\
 & = & - \sum_{n} {\rm Tr}\,\sum_{j=1}
{(-1)^{j}\over j+1} (eK_{n}^{(0)}A\!\!\!\slash)^{j+1}\label{a6}
\end{eqnarray}
where the intermediate phase factors cancel because of the gamma
matrix algebra, whereas the initial and the final phase factors cancel
because of cyclicity of the trace. It now follows that the parity
violating part of this action
\begin{equation}
\Gamma_{eff}^{{\rm PV} (higher)} [A, M] = {1\over 2}(\Gamma_{eff}^{(higher)}
[A,M] - \Gamma_{eff}^{(higher)} [A,-M]) = 0\label{a7}
\end{equation}
which follows because expression (\ref{a6}) is an even function of the fermion
mass. This shows that the parity violating part of the effective
action is at best linear in $\vec{A}$. However, as is clear from this
derivation, it says nothing about the parity conserving part of the
effective action, which, in general will contain higher powers of
$\vec{A}$. In fact, as we can see from Eq. (\ref{a6}), the parity
conserving part will have the quadratic term of the form
\begin{eqnarray}
\Gamma_{eff}^{PC (2)} & = &  {e^{2}\over 2}\sum_{n} {\rm Tr}\,
K_{n}^{(0)}A\!\!\!\slash K_{n}^{(0)}A\!\!\!\slash\nonumber\\
 & = &  {e^{2}\over 2}\sum_{n} {\rm Tr}\, {1\over p\!\!\!\slash +
\rho_{n}}A\!\!\!\slash {1\over p\!\!\!\slash +
\rho_{n}}A\!\!\!\slash\nonumber\\
 & = & - e^{2}\sum_{n}\int {\rm d}^{2}x\,{{\rm d}^{2}p\over
(2\pi)^{2}}\nonumber\\
 &  & \times\left({2p_{i}p_{j} + i(p_{i}\partial_{j}+p_{j}\partial_{i})
- \delta_{ij}(\rho_{n}^{2}+p_{k}(p_{k}+i\partial_{k}))\over
(\vec{p}^{2}+\rho_{n}^{2})((\vec{p}+i\vec{\nabla})^{2}+\rho_{n}^{2})}A_{j}\right)A_{i}
\nonumber\\
 & = & -{e^{2}\over 2\pi} \sum_{n} \int \rmd\alpha
\rmd^{2}x\,\alpha(1-\alpha)\,A_{i}{(\partial_{i}\partial_{j} -
\nabla^{2}\delta_{ij})\over
(-\alpha(1-\alpha)\nabla^{2} +\rho_{n}^{2})}\,A_{j}\nonumber\\
 & = & {e^{2}\over 2\pi} \sum_{n} \int \rmd\alpha
\rmd^{2}x\,\alpha(1-\alpha) B\,{1\over \rho_{n}^{2}
-\alpha(1-\alpha)\nabla^{2}}\,B,
\end{eqnarray}
where we have defined the magnetic field as
\be\label{magnetic}
B\equiv\epsilon_{ij}\partial_i A_j,\quad i,j=1,2.
\ee
and combined the denominators using the Feynman combination formula in
the intermediate steps. The sum over the discrete frequencies can be
done in a simple manner using the formulae
\begin{eqnarray}
\sum_{n=-\infty}^{\infty} {1\over \left({(2n+1)\pi\over
\beta}+{\theta\over \beta}\right)^{2} + \mu^{2}} & = & {\beta\over
4\mu}\left[\tanh {1\over 2}(\beta\mu - i\theta) + \tanh {1\over
2}(\beta\mu + i\theta)\right]\nonumber\\
 & = & {\beta\over \mu} {\partial\over
\partial\theta}\tan^{-1}\left(\tanh {\beta\mu\over 2} \tan{\theta\over
2}\right)\label{id} 
\end{eqnarray}
and leads to
\begin{eqnarray}
\Gamma_{eff}^{PC (2)} & = & {e^{2}\beta\over 8\pi} \int \rmd\alpha
\rmd^{2}x\,\alpha(1-\alpha)\,B {1\over
\sqrt{M^{2}-\alpha(1-\alpha)\nabla^{2}}}\nonumber\\
 &  & \times\left[\tanh {1\over
2}\left(\beta\sqrt{M^{2}-\alpha(1-\alpha)\nabla^{2}}-iea\right)\right.\nonumber\\
 &  & \;\left. +  \tanh {1\over
2}\left(\beta\sqrt{M^{2}-\alpha(1-\alpha)\nabla^{2}}+iea\right)\right]\,B
\nonumber\\ 
 & = & {e\beta\over 2\pi} \int \rmd\alpha \rmd^{2}x\,\alpha(1-\alpha) B
{1\over \sqrt{M^{2}-\alpha(1-\alpha)\nabla^{2}}}\nonumber\\
 &   & \times {\partial\over \partial a}\tan^{-1}\left(\tanh
{\beta\sqrt{M^{2}-\alpha(1-\alpha)\nabla^{2}}\over 2}\tan{ea\over
2}\right) B\label{a8}
\end{eqnarray}
This is completely in agreement with the results of \cite{13} and it
is clear  that this action is manifestly invariant under {\it large
gauge} transformations (see Eq. (\ref{a1})). (Namely, the arctan changes by a
constant under a {\it large gauge} transformation. However, the
quadratic effective action involves a derivative and, therefore, this
action is invariant under {\it large gauge} transformations.)

The term in the effective action, linear in $\vec{A}$, has to be
evaluated more carefully since this term, as it stands (see
Eq. (\ref{a3'})),  needs to be
regularized. It was suggested in \cite{13} to look alternately at the linear
term in the derivative of the effective action
\begin{equation}
{\partial\Gamma_{eff}^{(1)}\over \partial a} = {e^{2}\over \beta}
\sum_{n} {\rm Tr}\;K_{n}A\!\!\!\slash K_{n} \gamma_{0}
\end{equation}
This would correspond to making one subtraction. However, this
expression is still not fully regularized (it does not satisfy
cyclicity as can be easily checked) so that the effective action
linear in $\vec{A}$ was derived in \cite{13} in a limiting manner from
this.  Let us
note, however, that we are interested in the parity violating part of
the effective action. Thus, from Eq. (\ref{a2}), we obtain
\begin{eqnarray}
{\partial\Gamma_{eff}^{{\rm PV}}\over \partial a} & = & {1\over 2}
\left({\partial\Gamma_{eff} [A,M]\over \partial a} -
{\partial\Gamma_{eff} [A,-M]\over \partial a}\right)\nonumber\\
 & = & -{e\over 2\beta} \sum_{n} {\rm Tr}\left({1\over p\!\!\!\slash +
\gamma_{0}\tilde{\omega}_{n}+M+eA\!\!\!\slash} - {1\over p\!\!\!\slash +
\gamma_{0}\tilde{\omega}_{n}-M+eA\!\!\!\slash}\right)\gamma_{0}\nonumber\\
 & = &\!\!- {eM\over \beta}\! \sum_{n} {\rm Tr}\,{1\over p^{2}+
\tilde{\omega}_{n}^{2} + M^{2} + e(\vec{p}\cdot \vec{A}+\vec{A}\cdot
\vec{p}- i\gamma_{0}B) + e^{2}\vec{A}^{2}}\gamma_{0}\nonumber\\
 &   & 
\end{eqnarray}
The linear term (in $\vec{A}$) of this expression gives
\begin{equation}
{\partial\Gamma_{eff}^{{\rm PV} (1)}\over \partial a} =  {e^{2}M\over
\beta} \sum_{n} {\rm Tr}\,{1\over
p^{2}+\tilde{\omega}_{n}^{2}+M^{2}}
(\vec{p}\cdot\vec{A}+\vec{A}\cdot\vec{p}-i\gamma_{0}B) {1\over
p^{2}+\tilde{\omega}_{n}^{2}+M^{2}} \gamma_{0}
\end{equation}
This expression is well defined and satisfies the cyclicity
condition. Evaluating the Dirac trace gives (\lq\lq tr'' simply denotes trace
over a complete basis)
\begin{eqnarray}
{\partial\Gamma_{eff}^{{\rm PV} (1)}\over \partial a} & = & {2ie^{2}M\over \beta}
\sum_{n} {\rm tr}\,{1\over
(p^{2}+\tilde{\omega}_{n}^{2}+M^{2})^{2}}\,B\nonumber\\
 & = & {ie^{2}M\over 2\pi\beta} \sum_{n} \int \rmd^{2}x\,{1\over
\tilde{\omega}_{n}^{2}+M^{2}}\,B\nonumber\\
 & = & {ie^{2}\over 8\pi} \int \rmd^{2}x\,\left[\tanh{1\over
2}(\beta M-iea)+\tanh{1\over 2}(\beta M+iea)\right]B\nonumber\\
 & = & {ie\over 2\pi} {\partial\over \partial a} \int
\rmd^{2}x\,\tan^{-1}(\tanh{\beta M\over 2}\tan {ea\over 2}) B\label{a9}
\end{eqnarray}
This determines the parity violating effective action linear in $B$ which
precisely coincides with the action derived earlier \cite{6,7,13} and
for  future use, let
us define
\begin{equation}
\Gamma (a,M) = {e\over 2\pi} \,\tan^{-1}(\tanh{\beta
M\over 2} \tan{ea\over 2})\label{a10}
\end{equation} 
so that we can write 
\begin{equation}
\Gamma_{eff}^{{\rm PV} (1)} = i\int \rmd^{2}x\,B\,\Gamma (a,M)\label{a9'}
\end{equation}
(In general, of course, Eq. (\ref{a9}) determines the effective action
up to an additive constant. However, if we assume that the effective
action is normalized such that it vanishes when the external fields
vanish, then, the additive constant vanishes and Eq. (\ref{a9'}) gives
the parity violating part of the effective action linear in $\vec{A}$.)
Furthermore, as we have already shown, the parity violating part of
the effective action does not contain higher order terms in
$\vec{A}$. Consequently, this is the complete parity violating part of
the effective action in the particular background we have chosen,
consistent with the results in \cite{6}.

The particular gauge background, as we have argued earlier,
gives the leading terms in the static limit and,
consequently, this action would correspond to the leading order parity
violating effective action in that limit. We will next try
to extend these calculations to higher orders (in derivatives) and, in
the next section, we will describe the generalization in the momentum
space, before coming to the coordinate space calculations in the
following section.

\section{Momentum space calculations:}
Let us now consider momentum space amplitudes computed in terms
of the standard Feynman rules which can be derived from the Lagrangian 
density given by Eq. (\ref{lagra}), in the framework of finite
temperature field theory \cite{8,16,17}. (These amplitudes are the functional
coefficients, up to combinatoric factors and a temperature dependent
factor, of the  powers of the
external field $A_\mu$ in the series expansion of the effective action.)
Using the imaginary time formalism, 
one can express all the one-loop thermal amplitudes in terms of Bose
symmetric combinations  of the following basic $N$-point amplitudes
($\beta$ is the inverse temperature in units where the Boltzmann
constant is unity, as defined in the last section.)
\be\label{ba}
\begin{array}{lll}
{\cal A}_{\mu_1\cdots\mu_N}\left(\{k\};M\right) & = & -
\displaystyle{
\frac{e^N}{(2\pi)^2\beta}\sum_{n=-\infty}^{\infty}\int{\rmd}^2\vecp}
\\ & & \\ & &
\displaystyle{
\frac{{\cal N}_{\mu_1\cdots\mu_N}\left(p,\{k\};M\right)}
{\left(p^2-M^2\right)\cdots
\left[\left(p+k_{1(N-1)}\right)^2-M^2\right]}},
\end{array}
\ee
where $\{k\}\equiv k_1,\cdots,k_{N-1}$ represents the set of 
$N-1$ independent external 3-momenta, $k_{1i}\equiv k_1+k_2+\cdots +
k_i$,  and 
\be\label{trace1}
{\cal N}_{\mu_1\cdots\mu_N}=
\tr\left[\gamma_{\mu_1}
         \left(p\!\!\!\slash + k_1\!\!\!\!\!\slash\;+ M\right)
         \gamma_{\mu_2}
         \left(p\!\!\!\slash + k_{12}\!\!\!\!\!\!\!\slash\;\; + M\right)
         \cdots
         \gamma_{\mu_N}
         \left(p\!\!\!\slash + M\right)
    \right].
\ee
The external bosonic lines in Eq. (\ref{ba}) are such that the
zero component of its 3-momenta is quantized and purely imaginary 
(for instance ${k_1}_0={2i\pi\,l\over \beta}$, with $l=0,\pm 1, \pm 2,
\cdots$). Similarly, the zero component of the 3-momenta associated 
with a fermion loop is given by 
\be\label{omega}
p_0 = {i\,\pi\, \left(2 n + 1\right)\over \beta} \equiv i\,\omega_n;
\;\;\;\;\; n=0,\pm 1, \pm 2, \cdots.
\ee

Every thermal $N$-point amplitude is the sum of a parity
violating and a parity conserving part. In what follows, we will
concentrate only on the former, which can be written as (see also
Eq. (\ref{a7}))
\be\label{apv}
{\cal A}^{{\rm PV}}_{\mu_1\cdots\mu_N}\left(\{k\};M\right)=
\frac 1 2 
\left[{\cal A}_{\mu_1\cdots\mu_N}\left(\{k\};M\right) -
      {\cal A}_{\mu_1\cdots\mu_N}\left(\{k\};-M\right)\right],
\ee
Since the denominator in 
Eq. (\ref{ba}) is an even function of $M$, only the odd powers of $M$ from the
numerator ${\cal N}_{\mu_1\cdots\mu_N}$  in Eq. (\ref{apv}) will contribute to 
${\cal A}^{{\rm PV}}_{\mu_1\cdots\mu_N}$. Consequently, the parity violating
parts of  thermal amplitudes come only from those terms in Eq.
(\ref{trace1}) which involve the trace of an odd number of Dirac
gamma matrices.  

Expressing the two terms on the right hand side of Eq. (\ref{apv}) in
terms of the integral in Eq. (\ref{ba}) and performing the change of variable 
$p\rightarrow -p$, we can easily verify that
\be\label{ident1}
{\cal A}^{{\rm PV}}_{\mu_1\cdots\mu_N}\left(\{-k\};M\right)=
(-1)^{N+1} 
{\cal A}^{{\rm PV}}_{\mu_1\cdots\mu_N}\left(\{ k\};M\right).
\ee
This result confirms that the procedure of anti-symmetrization in the mass
gives a result which is in agreement with the usual concept of parity violation,
according to which the $N$-point amplitude is odd under the concomitant 
interchange of the sign of all external gauge fields as well as their 
respective momenta. 

\subsection{Gamma algebra}
In rotating the $2+1$ dimensional theory to imaginary time, one deals
with Euclidean Dirac matrices (see Eq. (\ref{a3''})) and  the trace of any
number of Euclidean  gamma matrices can be worked out in a
straightforward manner  by successive iteration of the
basic identity
\be\label{2gammas}
\gamma_{\mu_1}\gamma_{\mu_2}=-\delta_{{\mu_1}{\mu_2}} I
                             - \epsilon_{\mu_1\mu_2\lambda}\gamma_\lambda,
\ee
where $I$ is the two by two identity matrix and 
$\epsilon_{\mu_1\mu_2\mu_3}$ is the totally antisymmetric tensor.
Taking the trace of both sides of Eq. (\ref{2gammas}) we readily obtain
\be\label{tr2gammas}
\tr\,\gamma_{\mu_1}\gamma_{\mu_2} = - 2 \delta_{{\mu_1}{\mu_2}},
\ee
where we have used $\tr\,\gamma_\mu=0$. In the next step of iteration we
multiply both sides of Eq. (\ref{2gammas}) by $\gamma_{\mu_3}$, take
the trace and use Eq. (\ref{tr2gammas}). This simple calculation gives
\be\label{tr3gammas}
\tr\,\gamma_{\mu_1}\gamma_{\mu_2}\gamma_{\mu_3} =
2\, \epsilon_{\mu_1\mu_2\mu_3}.
\ee
When considering traces involving four or more gamma matrices we will
also have to take into account the following identity
\be\label{2eps}
\epsilon_{\mu_1\mu_2\mu_3} \epsilon_{\nu_1\nu_2\nu_3} = 
\det\left(\begin{array}{lcr}
          \delta_{\mu_1\nu_1} & \delta_{\mu_2\nu_1} & \delta_{\mu_3\nu_1} \\
          \delta_{\mu_1\nu_2} & \delta_{\mu_2\nu_2} & \delta_{\mu_3\nu_2} \\
          \delta_{\mu_1\nu_3} & \delta_{\mu_2\nu_3} & \delta_{\mu_3\nu_3} \\
          \end{array}\right)
\ee
For traces involving large numbers of gamma matrices,
we immediately realize that the algebra becomes increasingly more complicated.
For instance, the parity violating part of the four point function
(box diagram) involves the trace of up to seven gamma matrices.
However, the  pattern of iterations, which can be inferred from the first steps
described above, constitute a straightforward algorithm which
can be implemented, without difficulty, into a  computer
algebra program. Some of the calculations described in this section, 
which involve a relatively large number of gamma matrices, were, in fact, 
performed using computer algebra.

Of course, one can always draw some simple qualitative conclusions 
without having to perform the traces explicitly. For instance,
from the general pattern of the iterations described above, it is 
clear that any trace of an odd number of gamma matrices will necessarily
involve some combination of $\epsilon$-tensors (and $\delta$-tensors),
while  the trace of an
even number of gamma matrices is a combination of $\delta$-tensors.
As a result of this elementary property, the parity violating 
part of  amplitudes will always involve a single
$\epsilon$-tensor (higher number of tensors can always be reduced).

\subsection{The static limit}
Of course, we do not expect to be able to compute the amplitudes 
${\cal A}^{{\rm PV}}_{\mu_1\cdots\mu_N}$ for general arbitrary momenta
at finite temperature. This is because, at finite temperature,
amplitudes are non-analytic and, therefore, one can at best describe
them in some limit.  In what follows, we 
will calculate the thermal amplitudes in the static limit,  
${k_i}_0 = 0$, where {\it large gauge} invariance is known to
be an issue. In this limit, the parity violating part
of the basic amplitudes can be written as
\be\label{bastatic}
\begin{array}{lll}
{\cal A}^{{\rm static},\, {\rm PV}}_{\mu_1\cdots\mu_N}\left(\{k\};M\right) & = & 
\displaystyle{(-1)^{N+1}
\frac{e^N}{(2\pi)^2\beta}\sum_{n=-\infty}^{\infty}\int{\rmd}^2\vecp}
\\ & & \\ & &
\displaystyle{
\frac{{\cal N}^{{\rm static},\, {\rm PV}}_{\mu_1\cdots\mu_N}\left(p,\{k\};M\right)}
{\left(\vecp^{\;2}+M_\omega^2\right)\cdots
\left[\left(\vecp+\veck_{1(N-1)}\right)^2+M_\omega^2\right]}},
\end{array}
\ee
where $M_\omega^2\equiv \omega_n^2+M^2$, with $\omega_n$
given by Eq. (\ref{omega}), and
\be
{\cal N}^{{\rm static},\, {\rm PV}}_{\mu_1\cdots\mu_N}=\left.
\frac 1 2 \left[{\cal N}_{\mu_1\cdots\mu_N}\left(p,\{k\};M\right) -
                {\cal N}_{\mu_1\cdots\mu_N}\left(p,\{k\};-M\right)\right]
\right|_{{k_1}_0,\cdots,{k_{(N-1)}}_0 = 0}.
\ee

\subsection{Calculation of the integrals and sums using the derivative expansion}

To evaluate the  two dimensional integral in  Eq. (\ref{bastatic}), we
can use
the standard Feynman parameterization to combine  the $N$ denominators.
After performing appropriate shifts, the integration over 
$\vec p$ can be easily performed. However, since a closed
form expression for the Feynman parameter integrals is, in general, difficult for
arbitrary values of the external momenta $\veck_i$, we will have to
resort to some approximation. A simple but important
approximation is the limit $|\veck_i|\ll M_\omega$, in which case 
we can employ a derivative expansion (the term ``derivative'' originates from
the configuration space transformation \hbox{${k_i}\rightarrow -i\partial_{x_i}$}), 
expressing the result as a power series in the external momenta.
Then, at any given order of the derivative expansion, the Feynman 
parameters will show up in the integrand as a simple polynomial, which
can be easily evaluated. After the integration
over the Feynman parameters are done, we will be left with only a
sum over $n$ which
can be performed in a closed form. In fact, one can even write down in
advance all the possible structures of the sums involved at any given 
order of the derivative expansion by considering 
the canonical dimensions of the amplitudes,
$[{\cal A}_{\mu_1\cdots\mu_N}] = [{e^N\over \beta}][M^{(2-N)}]$ (note that in 
$2+1$ dimensions $\left[e^2\right]=\left[M\right]$). For instance,
in the order $2s+1$ ($s=0,1,\cdots$) of derivatives (momentum), 
the self-energy ($N=2$) will be given by
\be\label{seder}
\left.\Pi^{{\rm static,\, PV}}_{\mu_1\mu_2}\right|^{2s+1}=
- {e^2\over \beta} 
\left[
\left(\sum_{n=-\infty}^{\infty}\frac{1}{\left(M^2+\omega_n^2\right)^{s+1}}\right)
\,M{{\cal T}_1}^{(2s + 1)}_{\mu_1\mu_2}(k)\right],
\ee
while the corresponding order of the $N=4$ amplitude can be written
down as
\be\label{4ptder}
\begin{array}{lll}
\left.
{\cal A}^{{\rm static,\, PV}}_{\mu_1\mu_2\mu_3\mu_4}\right|^{2s+1} & = &
- \displaystyle{{e^4\over \beta}} 
\left[\displaystyle{
\left(\sum_{n=-\infty}^{\infty}\frac{1}{\left(M^2+\omega_n^2\right)^{s+2}}\right)
\,M  {{\cal T}_2}^{(2s + 1)}_{\mu_1\mu_2\mu_3\mu_4}(k_1,k_2,k_3)}\right.
\\ & & \\ & + & \left.\displaystyle{
\left(\sum_{n=-\infty}^{\infty}\frac{1}{\left(M^2+\omega_n^2\right)^{s+3}}\right)
\,M^3{{\cal T}_3}^{(2s + 1)}_{\mu_1\mu_2\mu_3\mu_4}(k_1,k_2,k_3)}
\right],
\end{array}
\ee
where the tensors ${{\cal T}}_r^{(2s + 1)}$ encode structures of degree 
$2s + 1$ in the respective momentum arguments. Similarly, at order
$2s$ of the derivatives, the $N=3$ and $N=5$ amplitudes can be written down as
\be
\begin{array}{lll}
\left.{\cal A}^{{\rm static,\, PV}}_{\mu_1\mu_2\mu_3}\right|^{2s} & = &
\displaystyle{{e^3\over \beta}}
\left[\displaystyle{
\left(\sum_{n=-\infty}^{\infty}\frac{1}{\left(M^2+\omega_n^2\right)^{s+1}}\right)
\,M  {{\cal T}_4}^{(2s)}_{\mu_1\mu_2\mu_3}(k_1,k_2)}\right.
\\ & & \\ & + & \left.\displaystyle{
\left(\sum_{n=-\infty}^{\infty}\frac{1}{\left(M^2+\omega_n^2\right)^{s+2}}\right)
\,M^3{{\cal T}_5}^{(2s)}_{\mu_1\mu_2\mu_3}(k_1,k_2)}
\right]
\end{array}
\ee
and
\be
\begin{array}{lll}
\left.{\cal A}^{{\rm static,\, PV}}_{\mu_1\mu_2\mu_3\mu_4\mu_5}\right|^{2s} & = &
\displaystyle{{e^5\over \beta}} 
\left[\displaystyle{
\left(\sum_{n=-\infty}^{\infty}\frac{1}{\left(M^2+\omega_n^2\right)^{s+2}}\right)
\,M  {{\cal T}_6}^{(2s)}_{\mu_1\mu_2\mu_3\mu_4\mu_5}(k_1,k_2,k_3,k_4)}\right.
\\ & & \\ & + & \displaystyle{
\left(\sum_{n=-\infty}^{\infty}\frac{1}{\left(M^2+\omega_n^2\right)^{s+3}}\right)
\,M^3{{\cal T}_7}^{(2s)}_{\mu_1\mu_2\mu_3\mu_4\mu_5}(k_1,k_2,k_3,k_4)}
\\ & & \\ & + & \left.\displaystyle{
\left(\sum_{n=-\infty}^{\infty}\frac{1}{\left(M^2+\omega_n^2\right)^{s+4}}\right)
\,M^5{{\cal T}_8}^{(2s)}_{\mu_1\mu_2\mu_3\mu_4\mu_5}(k_1,k_2,k_3,k_4)}
\right].
\end{array}
\ee
All the previous sums over $n$ have a closed form expression in terms
of derivatives of the basic sum (see Eq. (\ref{id}))
\be\label{sumn}
{\cal S}(\mu)\equiv\sum_{n=-\infty}^{\infty}\frac{1}{\mu^2+\omega_n^2} =
\frac{\beta}{2 \mu}\tanh{\left(\frac{\beta\mu}{2}\right)}.
\ee

\subsection{Explicit results up to the four point function}

In the Abelian theory, which we are considering, all the odd point
amplitudes vanish simply because of charge conjugation
invariance. Therefore, in the calculation of Feynman amplitudes in
this theory, we will concentrate only on the even point amplitudes.
Let us first show in some detail the derivation of the explicit results for the
self-energy ($N=2$) as well as the box diagram ($N=4$), which are implicitly
given in Eqs. (\ref{seder}) and (\ref{4ptder}), respectively. 
From Eq. (\ref{bastatic}), with $N=2$, the self-energy is given by
\be
\begin{array}{lll}
\Pi^{{\rm static},\, {\rm PV}}_{\mu_1\mu_2}\left(k\right) & \equiv &
{\cal A}^{{\rm static},\, {\rm PV}}_{\mu_1\mu_2}\left(k;M\right)
\\ & & \\ & = & \displaystyle{
- \frac{e^2}{(2\pi)^2\beta}\sum_{n=-\infty}^{\infty}\int{\rmd}^2\vecp
\frac{{\cal N}^{{\rm static},\, {\rm PV}}_{\mu_1\mu_2}\left(p,k;M\right)}
{\left(\vecp^{\;2}+M_\omega^2\right)
\left[\left(\vecp+\veck\right)^2+M_\omega^2\right]}}
\end{array}.
\ee
Using the Feynman combination formula, we can write
\be\label{sestpv}
\Pi^{{\rm static},\, {\rm PV}}_{\mu_1\mu_2}\left(k\right)=
-\frac{e^2}{(2\pi)^2\beta}\sum_{n=-\infty}^{\infty}\int_0^1{\rmd}\alpha\int{\rmd}^2\vecp
\frac{{\cal N}^{{\rm static},\, {\rm PV}}_{\mu_1\mu_2}
\left(p,k;M\right)}
{\left[(\vecp+\alpha\veck)^2 + \alpha(1-\alpha)\veck^{\;2} + M_\omega^2\right]^2}.
\ee
The parity violating part of the numerator involves a simple trace of 
only three gamma matrices. Using 
Eq. (\ref{tr3gammas}) one easily obtains
\be\label{numse}
{\cal N}^{{\rm static},\, {\rm PV}}_{\mu_1\mu_2}=
\left(2 M \, k_\alpha\epsilon_{\alpha,\mu_1,\mu_2}\right)^{{\rm static}}=
2 M \, k_j \epsilon_{j\mu_1\mu_2}.
\ee
Since we are in the static limit, namely $k_{0}=0$, either the 
index $\mu_{1}$ or $\mu_{2}$ has to be in the time direction. Choosing
$\mu_1=0$ and $\mu_2=i$, and noting that 
\be
\epsilon_{0ij}\equiv \epsilon_{ij}
\ee
we obtain
\be
\Pi^{{\rm static},\, {\rm PV}}_{0i}\left(k\right)=
- \epsilon_{ij}k_{j} \frac{2e^2 M}{(2\pi)^2\beta}
\sum_{n=-\infty}^{\infty}\int_0^1{\rmd}\alpha\int{\rmd}^2\vecp
\frac{1}
{\left[\vecp^{\;2} + \alpha(1-\alpha)\veck^{\;2} + M_\omega^2\right]^2},
\ee
where we have performed the shift $\vecp\rightarrow\vecp-\alpha\veck$.
The integration in $\vec p$ is now elementary, giving the result
\be\label{piexact1}
\Pi^{{\rm static},\, {\rm PV}}_{0i}\left(k\right)=
 - \epsilon_{ij}k_{j} \frac{e^2 M}{2\pi\beta}
\sum_{n=-\infty}^{\infty}\int_0^1{\rmd}\alpha
\frac{1}
{\alpha(1-\alpha)\veck^{\;2} + M^2 + \omega^2_n}.
\ee
We can now proceed in one of two ways, namely, either perform a
derivative  expansion, as
describe earlier, or perform the sum over $n$, using the formula
(\ref{sumn}) with \hbox{$\mu = \sqrt{\alpha(1-\alpha)\veck^{\;2} + M^2}$}.
Using the latter approach, we obtain
\be\label{piexact2}
\Pi^{{\rm static},\, {\rm PV}}_{0i}\left(k\right)=
 - \epsilon_{ij} k_{j} \frac{e^2 M}{4\pi}
\int_0^1{\rmd}\alpha
\frac{{ 
\tanh{\left(
\frac{\beta\sqrt{\alpha(1-\alpha)\veck^{\;2} + M^2}}{2}
\right)}}
}{\sqrt{\alpha(1-\alpha)\veck^{\;2} + M^2}}.
\ee
This expression shows that even
in the simplest case of the one-loop self-energy 
in the static limit, one cannot obtain a simple closed form expression.
Of course, the integration over the Feynman parameter can be performed order by
order using a derivative expansion of Eqs. (\ref{piexact1}) or (\ref{piexact2}).
As we have discussed  earlier in Eq. (\ref{seder}), each term of this
expansion is a function of degree $2s + 1$ ($s=0,1,\cdots$) in the
external momenta. It is clear from Eq. (\ref{piexact2})
that, at any $2s+1$ order, the polynomial in the Feynman parameter can be
systematically expressed in terms of Euler's beta function
${\rm B}$ which is defined as
\be\label{beta}
{\rm B}(s+1,s+1) = \int_0^1\rmd \alpha \alpha^s \left(1-\alpha\right)^s,
\ee
so that the expansion of the integrand in Eq. (\ref{piexact1}) in powers
of $(\veck)^2$ yields
\be\label{pider1}
\begin{array}{ll}
\Pi^{{\rm static},\, {\rm PV}}_{0i}\left(k\right)=
\displaystyle{
- \epsilon_{ij}k_{j} \frac{e^2 M}{2\pi}}
\displaystyle{
\sum_{s=0}^{\infty}}&
\displaystyle{
(-1)^s{\rm B}(s+1,s+1)(\veck^{\;2})^s}
\\ &\times
\displaystyle{\frac{(-1)^s}{s!}
\frac{\partial^s}{(\partial M^2)^s}
\left[\frac{1}{2 M}\tanh{\left(\frac{M}{2T}\right)}\right]}.
\end{array}
\ee
Using Eqs. (\ref{id}) and (\ref{a10}), we finally obtain
\be\label{pider2}
\begin{array}{ll}
\Pi^{{\rm static},\, {\rm PV}}_{0i}\left(k\right)=
\displaystyle{
 - \epsilon_{ij} k_{j}\,M}
\displaystyle{
\sum_{s=0}^{\infty}}\!\!&
\displaystyle{
\frac{{\rm B}(s+1,s+1)}{s!}} (\veck^{\;2})^s
\\ &\times
\displaystyle{
\frac{\partial^s}{(\partial M^2)^s}
\left(\frac{1}{M}\left.\frac{\partial}{\partial a}
\Gamma(a,M)\right|_{a=0}\right)}.
\end{array}
\ee
Equation (\ref{pider2}) gives the momentum space two point
amplitude which is obtained from the parity violating, quadratic
effective action by taking functional derivative with respect to
$A_0(\veck)$ and $A_i(-\veck)$.

Let us next consider the box diagram which is obtained from 
Eq. (\ref{bastatic}) with $N=4$. 
\be\label{box1}
\begin{array}{ll}
{\cal A}^{{\rm static},\, {\rm PV}}_{\mu_1\mu_2\mu_3\mu_4}
\left(k_1,k_2,k_3;M\right)  =  
\displaystyle{
- \frac{e^4}{(2\pi)^2\beta}\sum_{n=-\infty}^{\infty}\int{\rmd}^2\vecp}
& \\ & \\ 
\displaystyle{
\frac{{\cal N}^{{\rm static},\, {\rm PV}}_{\mu_1\mu_2\mu_3\mu_4}
\left(p,k_1,k_2,k_3;M\right)}
{\left(\vecp^{\;2}+M_\omega^2\right)
\left[\left(\vecp+\veck_{1  }\right)^2+M_\omega^2\right]
\left[\left(\vecp+\veck_{12 }\right)^2+M_\omega^2\right]
\left[\left(\vecp+\veck_{123}\right)^2+M_\omega^2\right]}} &,
\end{array}
\ee
From our experience with the previous example the self-energy,
we do not expect to obtain a closed form for Eq. (\ref{box1}) for
arbitrary values of $\veck$. Therefore, we will adopt right from
the beginning the derivative approximation $\veck_i\ll M_\omega$.
Furthermore, instead of trying to obtain the general term 
of the series, we will analyze separately each individual order up 
to the fifth order in the external momenta and consider the 
specific component $\mu_1=\mu_2=\mu_3=0, \mu_4=i$, which corresponds 
to the part of the effective action containing three $A_0$ fields 
and one magnetic field.

The parity violating numerator in Eq. (\ref{box1}) is an odd function
of the external momenta which can have degree one or three (this can
be easily verified from Eq. (\ref{trace1}) and the definition of
parity violating numerator as an antisymmetric function of $M$).
Making the external momenta equal to zero inside the
denominators and keeping only the linear contribution from the
numerator in Eq. (\ref{box1}), we obtain the following leading linear 
contribution
\be\label{box2}
\left.
{\cal A}^{{\rm static},\, {\rm PV}}_{000i}
\right|^{(1)}=  
- \frac{e^4}{(2\pi)^2\beta}\sum_{n=-\infty}^{\infty}\int{\rmd}^2\vecp\;
\frac{\left.{\cal N}^{{\rm static},\, {\rm PV}}_{000i}
\left(p,k_1,k_2,k_3;M\right)\right|^{(1)}}
{\left(\vecp^{\;2}+M^2+\omega_n^2\right)^4},
\ee
where
\be\label{numlin}
\left.{\cal N}^{{\rm static},\, {\rm PV}}_{000i}\right|^{(1)}=
2\,M \epsilon_{ij}\left[\left(3k_1+4k_2+3k_3\right)_j\omega_n^2-
\left(k_1+k_3\right)_{j}\left(\vecp^{\;2}+M^2\right)\right]
\ee
comes from the trace computation.
Substituting Eq. (\ref{numlin}) into Eq. (\ref{box2}) and
performing the the six permutations of the external momenta and indices
yields the following Bose symmetric expression for the box diagram
\be\label{boxlin1}
\left.\Pi^{{\rm static},\, {\rm PV}}_{000i}\right|^{(1)} = 
- 8\, \epsilon_{ij} {k_4}_j  
\frac{e^4 M}{(2\pi)^2\beta}\sum_{n=-\infty}^{\infty}\int{\rmd}^2\vecp\;
\frac{\vecp^{\;2}+M^2-5\omega_n^2}
{\left(\vecp^{\;2}+M^2+\omega_n^2\right)^4},
\ee
where we have used the momentum conservation $k_1+k_2+k_3=-k_4$.
Performing the integration over $\vecp$ in Eq. (\ref{boxlin1}), we
obtain
\be\label{boxlin2}
\left.\Pi^{{\rm static},\, {\rm PV}}_{000i}\right|^{(1)} = 
- 2\, \epsilon_{ij} {k_4}_j  
\frac{e^4 M}{2\pi\beta}\sum_{n=-\infty}^{\infty}
\left[\frac{4 M^2}
{\left(M^2+\omega_n^2\right)^3} -
\frac{3}
{\left(M^2+\omega_n^2\right)^2}\right].
\ee
Using Eq. (\ref{sumn}) we can perform the sum and express the result
in terms of derivatives of Eq. (\ref{a10}) in the following way
\be\label{boxlin3}
\left.\Pi^{{\rm static},\, {\rm PV}}_{000i}\right|^{(1)} = 
 \epsilon_{ij} {k_4}_j\;
\beta^{2}\left.\frac{\partial^3}{\partial a^3}\Gamma(a,M)\right|_{a=0}.
\ee

In order to obtain the higher order derivative contributions, 
we will have to take into account the external momenta
dependence inside the denominators of Eq. (\ref{box1}). Using the
Feynman combination formula we can write 
\be\label{box3}
\begin{array}{lll}
{\cal A}^{{\rm static},\, {\rm PV}}_{000i} &  =  &  
\displaystyle{
-\frac{6 e^4}{(2\pi)^2\beta}\sum_{n=-\infty}^{\infty}
\int_0^1{\rmd}\alpha_1
\int_0^{1-\alpha_1}{\rmd}\alpha_2\int_0^{1-\alpha_2}{\rmd}\alpha_3
\int{\rmd}^2\vecp}
\\ & & \\ & & 
\displaystyle{
\frac{{\cal N}^{{\rm static},\, {\rm PV}}_{000i}
\left(p_0,\vecp-\alpha_1 \veck_1-\alpha_2 \veck_{12}-\alpha_3
  \veck_{13},\veck_1,\veck_2,\veck_3;M\right)}
{\left(\vecp^{\;2}+M^2+\omega_n^2+K^2\right)^4}} ,
\end{array}
\ee
where 
$\veck_{12}\equiv \veck_1+\veck_2$, 
$\veck_{13}\equiv\veck_1+\veck_2+\veck_3$ and
\be
\begin{array}{lll}
K^2 & \equiv & \veck_{1 }^2\,\alpha_1\left(1-\alpha_1\right)+
               \veck_{12}^2\,\alpha_2\left(1-\alpha_2\right)+
               \veck_{13}^2\,\alpha_3\left(1-\alpha_3\right) \\ & & \\ 
& - & 2\left(\veck_1\cdot \veck_{12}\,\alpha_1\alpha_2+
        \veck_2\cdot \veck_{13}\,\alpha_2\alpha_3+
        \veck_1\cdot \veck_{13}\,\alpha_1\alpha_3\right)
\end{array}
\ee
Except for structures like
\[
M\, p_i p_j k_l \epsilon_{jl};\;\;\; {\rm or}\;\;\; 
M\, p_l p_l k_j \epsilon_{ij}
\]
which appear in the numerator
${\cal N}^{{\rm static},\, {\rm PV}}_{000i}$,
the ${\rmd}^2\vecp\;$ integration in Eq. (\ref{box3}) is as
straightforward as the ones that arose in the self-energy
calculation. In order to obtain a simple scalar integral we first
perform the elementary angular integrations with the help of 
\be
\int_0^{2\pi}{\rmd}\theta p_i\, p_j = \pi \vecp^{\;2} \delta_{ij}.
\ee
In this way, Eq. (\ref{box3}) leads to
\be\label{box4}
\begin{array}{lll}
{\cal A}^{{\rm static},\, {\rm PV}}_{000i} & = & 
\displaystyle{
-\frac{6 e^4}{2\pi\beta}\sum_{n=-\infty}^{\infty}
\int_0^1{\rmd}\alpha_1
\int_0^{1-\alpha_1}{\rmd}\alpha_2\int_0^{1-\alpha_2}{\rmd}\alpha_3}
\\ & & \\ & \times &\displaystyle{
\int_0^{\infty}p{\rmd}p}
\displaystyle{
\frac{n_i^{(1)}\vecp^{\;2}+N_i^{(1)}+N_i^{(3)}}
{\left(\vecp^{\;2}+M^2+\omega_n^2+K^2\right)^4}}.
\end{array}
\ee
The compact notation in the numerator of Eq. (\ref{box4}) means that
$n_i^{(1)}$ and $N_i^{(1)}$ are of first order in the external
momenta, while $N_i^{(3)}$ is of third order in the external momenta.
(Of course, the algebra has become very much involved by now. Just to 
give an idea of how involved it is, the numerator in Eq. (\ref{box4})
contains $242$ terms.) Performing the integration in ${\rmd}p$ and 
expanding the result up to fifth order in the external momenta yields
the following third and fifth order expressions
\be\label{box3rd}
\begin{array}{lll}
\left.
{\cal A}^{{\rm static},\, {\rm PV}}_{000i}\right|^{(3)} & = & 
-\displaystyle{
\frac{e^4}{2\pi\beta}\sum_{n=-\infty}^{\infty}
\int_0^1{\rmd}\alpha_1
\int_0^{1-\alpha_1}{\rmd}\alpha_2\int_0^{1-\alpha_2}{\rmd}\alpha_3}
\\ & & \\ & \times &
\displaystyle{\left[
\frac{N_i^{(3)}-K^2n_{i}^{(1)}}{(M^2+\omega_n^2)^3}-
3 \frac{K^2 N_{i}^{(1)}}{(M^2+\omega_n^2)^4}\right]}
\end{array}
\ee
and
\be\label{box5th}
\begin{array}{lll}
\left.
{\cal A}^{{\rm static},\, {\rm PV}}_{000i}\right|^{(5)} & = & 
-\displaystyle{
\frac{e^4}{2\pi\beta}\sum_{n=-\infty}^{\infty}
\int_0^1{\rmd}\alpha_1
\int_0^{1-\alpha_1}{\rmd}\alpha_2\int_0^{1-\alpha_2}{\rmd}\alpha_3}
\\ & & \\ & \times &
\displaystyle{\left[
\frac{6 K^4 N_{i}^{(1)}}{(M^2+\omega_n^2)^5}+
\frac 3 2 \frac{K^4 n_{i}^{(1)}-2 K^2 N_{i}^{(3)}}{(M^2+\omega_n^2)^4}\right]}.
\end{array}
\ee
The parametric integrals in the above expressions are very involved,
but straightforward, since there are only powers of the Feynman parameters.
As in the previous cases, the sum over discrete energy can also 
be performed using Eq. (\ref{sumn}) and the result can be expressed in
terms of derivatives of $\Gamma(a,M)$ defined in Eq. (\ref{a10}).
The complete four photon amplitude is then obtained adding the six
permutations of external momenta and indices.

Of course, the final result must preserve the small gauge invariance, 
being proportional to $\epsilon_{ij} {k_4}_j$, like the leading order 
result given by Eq. (\ref{boxlin3}), so that the contraction with 
${k_4}_i$ gives zero (this is a consequence of the invariance under 
a small gauge transformation
$\vec A(k_4)\rightarrow \vec A(k_4) + \veck_4$  in the momentum space).
However, at this higher order, our explicit calculation shows that the 
small gauge invariance will only be explicitly manifest, when we make
use of some identities involving the 2-dimensional vectors. A simple 
example is the Jacobi identity 
\be\label{jacobi}
\left({k_1}_l\,{k_2}_m\,{k_3}_i +
      {k_2}_l\,{k_3}_m\,{k_1}_i +
      {k_3}_l\,{k_1}_m\,{k_2}_i\right)\epsilon_{lm} = 0.
\ee
The emergence of these identities is, in fact, expected, because
the very nature of the sub-leading contributions (higher powers of the
external momenta) leaves room to write the two-dimensional structures 
involving $\epsilon_{ij}$ and the vectors $\veck_1$,  $\veck_2$ and 
$\veck_3$ in many equivalent ways. Our strategy to single out the unique
gauge invariant form, was to decompose each vector in a
two-dimensional basis and verify (by brute force, using
the computer) that the unique function of the components is indeed 
gauge invariant. Then, from the expressions in terms of components, we
were able to identify the two-dimensional scalar functions which multiplies 
$\epsilon_{ij} {k_4}_j$. This leads to the following results
\newpage
\be\label{box3rd1}
\begin{array}{lll}
\left.\Pi^{{\rm static},\, {\rm PV}}_{000i}\right|^{(3)} & = &
\displaystyle{ 
\frac{e^4}{3\pi} \epsilon_{ij} {k_4}_j
\left(\veck_1^{\;2}+\veck_2^{\;2}+\veck_3^{\;2}+\veck_1\cdot \veck_2+
\veck_2\cdot \veck_3+\veck_2\cdot \veck_3\right)}
\\ & & \\ & \times &
\displaystyle{{M\over \beta}
\sum_{n=-\infty}^{\infty}\frac{M^2-5\omega_n^2}{(M^2+\omega_n^2)^4}}
\\ & & \\ & = &\displaystyle{
\frac{\epsilon_{ij} {k_4}_j}{6}
\left(\veck_1^{\;2}+\veck_2^{\;2}+\veck_3^{\;2}+\veck_1\cdot \veck_2
+\veck_2\cdot \veck_3+\veck_2\cdot \veck_3\right)}
\\ & & \\ & \times &\displaystyle{M\,\beta^{2}
\frac{\partial}{\partial M^2}\left[\frac 1 M
\frac{\partial^3}{\partial a^3}\Gamma(a,M)\right]_{a=0}}
\end{array}
\ee
and
\be\label{box5th1}
\begin{array}{lll}
\left.\Pi^{{\rm static},\, {\rm PV}}_{000i}\right|^{(5)} 
& = &
\displaystyle{\frac{e^4 M}{30\pi\beta}\epsilon_{ij} {k_4}_j\left\{
\left(\veck_1^{\;2} \, \veck_2\cdot \veck_3 + 
\veck_1\cdot \veck_2 \,\veck_2\cdot \veck_3\right)\right.}
\displaystyle{\sum_{n=-\infty}^{\infty}\frac{2}{(M^2+\omega_n^2)^4}}
\\ & & \\ & - &\!\!\!\!\!\!
\left[\veck_1^{\;2}\left(3\veck_1^{\;2}+6\veck_1\cdot \veck_2 + 
6\veck_1\cdot \veck_3 + 5\veck_2^{\;2} +
5\veck_2\cdot\veck_3\right)\right. 
\\ & & \\ & + &\!\!\!\!\!\! 
\left.\left.4(\veck_1\cdot \veck_2)^2 + 
6 \veck_1\cdot \veck_2 \,\veck_2\cdot \veck_3 \right]
\displaystyle{
\sum_{n=-\infty}^{\infty}\frac{7\omega_n^2-M^2}{(M^2+\omega_n^2)^5}}\right\}
\\ & & \\ & + &
{\rm two\;cyclic\;permutations\; of\;} \veck_1,\;\veck_2,\; {\rm and} \; \veck_3
\\ & & \\ & = &
\displaystyle{ 
- \frac{M}{60}\epsilon_{ij} {k_4}_j\left\{
\left(\veck_1^{\;2} \, \veck_2\cdot \veck_3 + 
\veck_1\cdot \veck_2 \,\veck_2\cdot \veck_3\right)\right.}
\\ & & \\ & \;\;\;\;\;\;\times\!\!\!\! &\displaystyle{
\frac{4 e^2}{3} \frac{\partial^3}{(\partial M^2)^3}\left(
\frac 1 M \frac{\partial\Gamma}{\partial a}\right)_{a=0}}
\\ & & \\ & + &\!\!\!\!\!\!
\left[\displaystyle{
\veck_1^{\;2}\left(3\veck_1^{\;2}+6\veck_1\cdot \veck_2 + 
6\veck_1\cdot \veck_3 + 5\veck_2^{\;2} + 5\veck_2\cdot\veck_3\right)}\right.
\\ & & \\ & + &\!\!\!\!\!\! 
\left. \left. 4(\veck_1\cdot \veck_2)^2 + 
6 \veck_1\cdot \veck_2 \,\veck_2\cdot \veck_3 \right]
\displaystyle{
\frac {\beta^{2}} {3} \frac{\partial^2}{(\partial M^2)^2}\left(
\frac 1 M \frac{\partial^3\Gamma}{\partial a^3}\right)_{a=0}}
\right\}
\\ & & \\ & + &
{\rm two\;cyclic\;permutations\; of\;} 
\veck_1,\;\veck_2,\; {\rm and}\; \veck_3.
\end{array}
\ee
\newpage
This gives an idea of the nature of calculations in the momentum
space. With each higher order, the calculations become increasingly
more difficult. Besides, once the amplitudes are obtained, the
construction of the effective action from these is a nontrivial
operation, particularly at higher orders where several tensor
structures may be present. This forces us to look beyond the momentum
space calculations which we will do in the next section. However, it
is worth emphasizing at this point that although the momentum space
calculations have yielded basic Feynman amplitudes, one can easily
obtain from them corresponding amplitudes with any number of zero
momentum $A_{0}$ fields, simply by considering the fermion propagator
in the background of a constant $A_{0}$ field, as will become clear in
the next section.

\section{Derivative expansion at higher orders:}

In trying to determine the higher order terms (derivatives) in the
static limit, we let the $A_{0}$ field depend on space as well (in
contrast to the discussion in section {\bf 2}, Eq. (\ref{a1'})) and
make the decomposition
\begin{equation}\label{c1'}
A_{0}(t,\vec{x}) = \bar{A}_{0}(t) + \hat{A}_{0}(\vec{x}),\qquad \int
\rmd^{2}x\,\hat{A}_{0}(\vec{x}) = 0
\end{equation}
Namely, we have separated out the zero mode of the space dependent part
into the first term, which can always be done using a box
normalization. Once again, by a suitable gauge transformation (see
Eq. (\ref{gauge})), the
gauge fields can be brought to the form
\begin{equation}
A_{0}(t,\vec{x}) \rightarrow {a\over \beta} + \hat{A}_{0}(\vec{x}),\qquad \vec{A} =
\vec{A}(\vec{x})\label{c1}
\end{equation}
With such a separation, we have also separated the behaviour of the
fields under a {\it small} and a {\it large} gauge
transformation. Namely, under a {\it large gauge} transformation only
$a$ transforms as
\begin{equation}
a\rightarrow a + {2\pi n\over e}\label{c2}
\end{equation}
while under a {\it small gauge} transformation only $\vec{A}$
transforms as ($\hat{A}_{0}$ does not transform under a {\it small
gauge} transformation in the static limit, since we have already used
this freedom to bring $A_{0}$ to the form in Eq. (\ref{c1}).),
\begin{equation}
\vec{A}\rightarrow \vec{A}+\vec{\nabla}\epsilon
\end{equation}
In this case, the effective action (see Eq. (\ref{a3'})) takes the form
\begin{equation}
\Gamma_{eff} [A,M] = - \sum_{n}{\rm Tr}\,\ln\left(1+
(p\!\!\!\slash + \gamma_{0}(\tilde{\omega}_{n}+e\hat{A}_{0}) +
M)^{-1}\,(eA\!\!\!\slash)\right)\label{c3}
\end{equation}
The linear term in $\vec{A}$ has the simple form
\begin{equation}
\Gamma_{eff}^{(1)} [\hat{A}_{0},M] = - e \sum_{n} {\rm Tr}\,{1\over
p\!\!\!\slash + \gamma_{0}(\tilde{\omega}_{n}+e\hat{A}_{0}) +
M}\,A\!\!\!\slash\label{c3'} 
\end{equation}
It is clear now that if we expand the denominator in powers of
$\hat{A}_{0}$ and carry out the trace, we will obtain all the higher
derivative corrections to the effective action in the static
limit. However, it is also clear that the expansion would bring out
more and more factors of Dirac matrices in the numerator so that
calculations will become increasingly difficult as we go to higher
orders. Thus, we look for an alternate method for obtaining the
result.

Let us note that we are really interested in the parity violating part
of the effective action, which is obtained as
\begin{equation}
\Gamma_{eff}^{{\rm PV} (1)} = {1\over 2} (\Gamma_{eff}^{(1)} [\hat{A}_{0},M] -
\Gamma_{eff}^{(1)} [\hat{A}_{0},-M])\label{c4}
\end{equation}
Furthermore, let us also note the identity that
\begin{eqnarray}
 &  & {1\over 2}\left({1\over p\!\!\!\slash +
 \gamma_{0}(\tilde{\omega}_{n}+e\hat{A}_{0}) +M} - {1\over
 p\!\!\!\slash + \gamma_{0}(\tilde{\omega}_{n}+e\hat{A}_{0})
 -M}\right)\nonumber\\
 &  &\;\;\; = {M\over p^{2}+\tilde{\omega}_{n}^{2}+M^{2}+L}\label{c5}
\end{eqnarray}
where
\begin{equation}
L = 2e\tilde{\omega}_{n}\hat{A}_{0} -
ie\gamma_{0}(\partial\!\!\!\slash\hat{A}_{0}) +
e^{2}\hat{A}_{0}^{2}\label{c6} 
\end{equation}
contains all the field dependent terms and has a much simpler Dirac
matrix structure. Using this, we can write
\begin{equation}
\Gamma_{eff}^{{\rm PV} (1)} = - eM \sum_{n} {\rm Tr}\,{1\over
p^{2}+\tilde{\omega}_{n}^{2}+M^{2}+L}\,A\!\!\!\slash
\end{equation}
The denominator can now be expanded and the effective action can be
calculated for any number of $\hat{A}_{0}$ fields in a simple and
systematic manner.

As an  example, let us note that the part of the parity violating
action containing one $\hat{A}_{0}$ field in addition to the $B$ field
arises as 
\begin{eqnarray}
\left(\Gamma_{eff}^{{\rm PV} (1)}\right)^{(1)} & = &   eM \sum_{n} {\rm
Tr}\,{1\over
p^{2}+\tilde{\omega}_{n}^{2}+M^{2}}\,
(-ie\gamma_{0}(\partial\!\!\!\slash\hat{A}_{0})
{1\over p^{2}+\tilde{\omega}_{n}^{2}+M^{2}}A\!\!\!\slash\nonumber\\
 & = & - 2ie^{2}M \sum_{n} {\rm tr} {1\over
p^{2}+\tilde{\omega}_{n}^{2}+M^{2}} {1\over
(\vec{p}+i\vec{\nabla})^{2}+\tilde{\omega}_{n}^{2}+M^{2}}\,
(\partial_{i}\hat{A}_{0})\epsilon_{ij}A_{j}\nonumber\\ 
 & = &  2ie^{2}M \sum_{n}\int \rmd^{2}x{\rmd^{2}p\over
(2\pi)^{2}}\,{1\over p^{2}+\tilde{\omega}_{n}^{2}+M^{2}}\nonumber\\
 &   & \qquad\qquad\times {1\over
(\vec{p}+i\vec{\nabla})^{2}+\tilde{\omega}_{n}^{2}+M^{2}}\,\hat{A}_{0}B
\end{eqnarray}  
Here, the derivatives act only on $\hat{A}_{0}$ and not on $B$.
The momentum integral can be evaluated by combining the
denominators using the Feynman combination formula. Even the sum over
the discrete frequency modes can also be exactly evaluated (see
Eq. (\ref{id})) and the
parity violating effective action containing one $\hat{A}_{0}$ field
in addition to the $B$ field has the form
\begin{eqnarray}
\left(\Gamma_{eff}^{{\rm PV} (1)}\right)^{(1)} & = & {i e^{2}M\beta\over 8\pi} \int
\rmd^{2}x\rmd\alpha\,B\,{1\over
\sqrt{M^{2}-\alpha(1-\alpha)\nabla^{2}}}\nonumber\\
 &  &\times\left[\tanh {1\over
2}\left(\beta\sqrt{M^{2}-\alpha(1-\alpha)\nabla^{2}}-iea\right)\right.\nonumber\\
 &  & \;\;\left. +
\tanh {1\over
2}\left(\beta\sqrt{M^{2}-\alpha(1-\alpha)\nabla^{2}}+iea\right)\right]\hat{A}_{0}
\end{eqnarray}
This is an exact, closed form expression which can also be expanded in
powers of derivatives and takes the form
\begin{eqnarray}\label{A0B}
\left(\Gamma_{eff}^{{\rm PV} (1)}\right)^{(1)} & = & {i e^{2}M\beta\over 8\pi}
\sum_{s=0}^{\infty}\int \rmd^{2}x\,{B(s+1,s+1)\over s!}\,B
\left((-\nabla^{2})^{s}\hat{A}_{0}\right)\nonumber\\
 &  &\!\! \times {\partial^{s}\over
(\partial M^{2})^{s}}\left[{1\over M}(\tanh {1\over 2}(\beta M-iea) + \tanh
{1\over 2}(\beta M+iea)\right]\nonumber\\
 & = & iM\beta \sum_{s=0}^{\infty}\int \rmd^{2}x\,
{B(s+1,s+1)\over s!}\,B((-\nabla^{2})^{s}\hat{A}_{0}) \nonumber\\
 &  & \qquad\times {\partial^{s}\over (\partial M^{2})^{s}}
\left({1\over M}\frac{\partial}{\partial a}\Gamma (a,M)\right)
\end{eqnarray}
which can be compared with the result from the momentum space
result given by Eq. (\ref{pider2}) (recall that the coefficients of
the  momentum space amplitudes are related to those of the real space
amplitudes by a factor of ${i\over \beta}$).
Let us write out the first few terms explicitly which have the forms
\begin{eqnarray}
\left(\Gamma_{eff}^{{\rm PV} (1)}\right)^{(1)} & = & i\beta\int
\rmd^{2}x\,B\left[\hat{A}_{0}{\partial\Gamma\over \partial a} - {M\over 6}
(\nabla^{2}\hat{A}_{0}) {\partial\over \partial M^{2}}
\left(\frac 1 M {\partial \Gamma\over\partial a}\right)\right.\nonumber\\
 &  & \left. +{M\over 60} (\nabla^{4}\hat{A}_{0}) {\partial^{2}\over
(\partial M^{2})^{2}}\left(\frac 1 M
{\partial \Gamma\over\partial a}\right)+
\cdots\right]\label{c7} 
\end{eqnarray}
We note here that this effective action will give the amplitude of the
type $\hat{A}_{0}$-$B$ with any number of $a$ insertions which can be
thought of as zero momentum $A_{0}$ fields.

Without going into details, let us simply note here that, the parity
violating effective action containing two $\hat{A}_{0}$ fields, in
addition to the $B$ field (and, of course, any number of $a$ fields),
can also  be evaluated in a similar manner
and has the following form
\begin{eqnarray}
\left(\Gamma_{eff}^{{\rm PV} (1)}\right)^{(2)} & = & - 4ie^{3}M 
\sum_{n} {\rm tr}\,{\tilde{\omega}_{n}\over
(p^{2}+\tilde{\omega}_{n}^{2}+M^{2})((\vec{p}+i\vec{\nabla}_{1})^{2}+
\tilde{\omega}_{n}^2+M^{2})}\nonumber\\ 
 &  & \times
{1\over((\vec{p}+i\vec{\nabla}_{1}+i\vec{\nabla}_{2})^{2}+
\tilde{\omega}_{n}^{2}+M^{2})} 
\hat{A}_{0}^{(1)}\hat{A}_{0}^{(2)}B
\end{eqnarray}
Here, we have put indices on the derivatives as well as the $\hat{A}_{0}$
fields to indicate the action of these operators. The momentum
integral as well as the sum over the discrete frequencies can be
carried out in this case also and the final form can be obtained in a
closed form. However, let us make a power series expansion in the
derivatives and write down explicitly the first few terms
\begin{eqnarray}
\left(\Gamma_{eff}^{{\rm PV} (1)}\right)^{(2)} & = & i\beta^{2}\int
\rmd^{2}x\,B\left[{1\over 2!}\hat{A}_{0}^{2} 
{\partial^{2}\Gamma\over \partial a^{2}}\right.\nonumber\\
 &  &\!\!\! - {M\over 12} (2(\nabla^{2}\hat{A}_{0})\hat{A}_{0} +
(\vec{\nabla}\hat{A}_{0})\cdot (\vec{\nabla}\hat{A}_{0}))
{\partial\over \partial M^{2}}
\left(\frac 1 M{\partial^2\Gamma\over\partial a^2}\right)\nonumber\\
 &  &\!\!\! + {M\over 60} (2\hat{A}_{0}(\nabla^{4}\hat{A}_{0}) +
4(\nabla^{2}\partial_{i}\hat{A}_{0})(\partial_{i}\hat{A}_{0}) +
{4\over 3} (\partial_{i}\partial_{j}\hat{A}_{0})^{2} + {5\over 3}
(\nabla^{2}\hat{A}_{0})^{2})\nonumber\\
 &  & \left.\; \times {\partial^{2}\over (\partial M^{2})^{2}} 
\left(\frac 1 M {\partial^2\Gamma\over \partial a^2}\right)+
\cdots\right]\label{c8}
\end{eqnarray}

Calculations become algebraically more tedious as we go to higher
orders. For example, the parity violating part of the effective action
containing three $\hat{A}_{0}$ fields in addition to the $B$ field
(and any number of $a$ fields) can also be
evaluated and has the form (before simplification)
\begin{eqnarray}
\left(\Gamma_{eff}^{{\rm PV} (1)}\right)^{(3)} & = &  2ie^{4}M
\sum_{n} {\rm tr}\nonumber\\
 &  & \times\left[{1\over p^{2}+\tilde{\omega}_{n}^{2}+M^{2}}\hat{A}_{0}^{2}
{1\over p^{2}+\tilde{\omega}_{n}^{2}+M^{2}}(\partial_{i}\hat{A}_{0})
{1\over p^{2}+\tilde{\omega}_{n}^{2}+M^{2}}
\epsilon_{ij}A_{j}\right.\nonumber\\
 &  & \;\; +{1\over p^{2}+\tilde{\omega}_{n}^{2}+M^{2}}
(\partial_{i}\hat{A}_{0}) {1\over p^{2}+\tilde{\omega}_{n}^{2}+M^{2}}
\hat{A}_{0}^{2} {1\over p^{2}+\tilde{\omega}_{n}^{2}+M^{2}}
\epsilon_{ij}A_{j}\nonumber\\
 &  & \;\; +{4\tilde{\omega}_{n}^{2}\over
p^{2}+\tilde{\omega}_{n}^{2}+M^{2}} \hat{A}_{0} {1\over
p^{2}+\tilde{\omega}_{n}^{2}+M^{2}} \hat{A}_{0} {1\over
p^{2}+\tilde{\omega}_{n}^{2}+M^{2}} \hat{A}_{0}\nonumber\\
 &  & \times {1\over
p^{2}+\tilde{\omega}_{n}^{2}+M^{2}} B\nonumber\\
 &  & \;\; - {1\over p^{2}+\tilde{\omega}_{n}^{2}+M^{2}}
(\partial_{k}\hat{A}_{0}) {1\over p^{2}+\tilde{\omega}_{n}^{2}+M^{2}}
(\partial_{k}\hat{A}_{0}) {1\over
p^{2}+\tilde{\omega}_{n}^{2}+M^{2}}\nonumber\\
 &  & \;\;\times
(\partial_{i}\hat{A}_{0}) {1\over p^{2}+\tilde{\omega}_{n}^{2}+M^{2}}
\epsilon_{ij}A_{j}\nonumber\\
 &  & \;\; - {\epsilon_{ij}\over p^{2}+\tilde{\omega}_{n}^{2}+M^{2}}
(\partial_{i}\hat{A}_{0}) {1\over
p^{2}+\tilde{\omega}_{n}^{2}+M^{2}} (\partial_{j}\hat{A}_{0}) {1\over
p^{2}+\tilde{\omega}_{n}^{2}+M^{2}}\nonumber\\
 &  & \;\;\times\left. (\partial_{k}\hat{A}_{0}) {1\over
p^{2}+\tilde{\omega}_{n}^{2}+M^{2}} A_{k}\right]\label{c8'}
\end{eqnarray}
It is interesting to note that the expression above does not look
manifestly invariant under {\it small} gauge transformations. We
will give a proof of gauge invariance later. For
the present, let us simply note that if we were to evaluate this
expression in powers of  derivatives, the leading order term, which
will be linear in the derivatives has the form
\begin{eqnarray}
 & = & -{ie^{4}M\over 3\pi^{2}} \sum_{n} \int \rmd^{2}x\rmd^{2}p\, \left[{1\over
 (p^{2}+\tilde{\omega}_{n}^{2}+M^{2})^{3}} -
 {6\tilde{\omega}_{n}^{2}\over
 (p^{2}+\tilde{\omega}_{n}^{2}+M^{2})^{4}}\right]B \hat{A}_{0}^{3}
 \nonumber\\
 & = & -{ie^{4}M\over 6\pi} \sum_{n} \int \rmd^{2}x\left[{1\over
 (\tilde{\omega}_{n}^{2}+ M^{2})^{2}} - {4\tilde{\omega}_{n}^{2}\over
 (\tilde{\omega}_{n}^{2}+ M^{2})^{3}}\right] B\hat{A}_{0}^{3}\nonumber\\
 & = & {ie^{2}M\beta^{2}\over 12\pi} \sum_{n} \int
 \rmd^{2}x\,{\partial^{2}\over \partial a^{2}}\left({1\over
 \tilde{\omega}_{n}^{2}+M^{2}}\right) B\hat{A}_{0}^{3}\nonumber\\
 & = &  i\beta^{3}\int \rmd^{2}x\,{1\over 3!}\,B\hat{A}_{0}^{3}\,
 {\partial^{3}\Gamma\over \partial a^{3}}\label{c9}
\end{eqnarray}
This can again be compared with Eq. (\ref{boxlin3}).

The term cubic (only odd powers of derivatives arise) in the derivatives
has the form
\begin{eqnarray}\label{c10}
  \!\!\!\!\! \!\!\!\!\!\!\!\!\!\! \!\!\!\!\!
& & 
 \!\!\!\!\! \!\!\!\!\!\!\!\!\!\! \!\!\!\!\! =
{ie^{4}M\over 2\pi^{2}} \sum_{n} \int
 \rmd^{2}x\rmd^{2}p\,\left({8\tilde{\omega}_{n}^{2}\over
 (p^{2}+\tilde{\omega}_{n}^{2}+M^{2})^{5}} - {1\over
 (p^{2}+\tilde{\omega}_{n}^{2}+M^{2})^{4}}\right)\nonumber\\
 &  & \;\;\times((\nabla^{2}\hat{A}_{0})
\hat{A}_{0} + (\vec{\nabla}\hat{A}_{0})\cdot
 (\vec{\nabla}\hat{A}_{0})) \hat{A}_{0} B\nonumber\\
 \!\!\!\!\! \!\!\!\!\!\!\!\!\!\! \!\!\!\!\!
 & & \!\!\!\!\! \!\!\!\!\!\!\!\!\!\!\!\!\!\!\!\!  =
-{iM\beta^{3}\over 12} \int
 \rmd^{2}x\,B\hat{A}_{0}((\nabla^{2}\hat{A}_{0})\hat{A}_{0}
 +(\vec{\nabla}\hat{A}_{0})\cdot
 (\vec{\nabla}\hat{A}_{0}))\,{\partial\over \partial M^{2}}
\left(\frac 1 M \frac{\partial^3\Gamma}{\partial a^3}\!\right)
\end{eqnarray}
which agrees with Eq. (\ref{box3rd1}). Finally, the term fifth order
in the  derivatives has the form
\begin{eqnarray}
 & = & -\sum_{n}{ie^{4}M\over 20\pi (\tilde{\omega}_{n}^{2}+M^{2})^{5}}
 \int
 \rmd^{2}x\left[(7\tilde{\omega}_{n}^{2}-M^{2})\left(\hat{A}_{0}^{2}((\nabla^{2})^{2}
\hat{A}_{0})B
 \right.\right.\nonumber\\
 &  & \left. +
 4\hat{A}_{0}(\partial_{k}\hat{A}_{0})(\partial_{k}\nabla^{2}\hat{A}_{0})B
 +{5\over 3}
 \hat{A}_{0}(\nabla^{2}\hat{A}_{0})^{2} B + {4\over 3}
 \hat{A}_{0}(\partial_{k}\partial_{l}\hat{A}_{0})^{2}
 B\right)\nonumber\\
 &  &\;\;+(11\tilde{\omega}_{n}^{2}-{7M^{2}\over
 3})(\partial_{k}\hat{A}_{0})^{2} (\nabla^{2}\hat{A}_{0}) B\nonumber\\
 &  &\;\;\left. +
 ({40\tilde{\omega}_{n}^{2}\over 3} - {8M^{2}\over
 3})(\partial_{k}\hat{A}_{0})(\partial_{l}\hat{A}_{0})
 (\partial_{k}\partial_{l}\hat{A}_{0}) B\right]\nonumber\\
 & = & -{iM\over 120} \int
 \rmd^{2}x\,B\left[\beta^{3}\left(\hat{A}_{0}^{2}(\nabla^{4}\hat{A}_{0}) +
 4\hat{A}_{0}(\partial_{i}\hat{A}_{0})
 (\nabla^{2}\partial_{i}\hat{A}_{0}) + {5\over 3}
 \hat{A}_{0}(\nabla^{2}\hat{A}_{0})^{2}\right.\right.\nonumber\\
 &   & \left.\!\!\!\!\!\!\!\!+ {4\over 3}
 \hat{A}_{0}(\partial_{i}\partial_{j}\hat{A}_{0})^{2} + {5\over 3}
 (\partial_{i}\hat{A}_{0})^{2} (\nabla^{2}\hat{A}_{0}) + 2
 (\partial_{i}\hat{A}_{0})(\partial_{j}\hat{A}_{0})(\partial_{i}\partial_{j}
\hat{A}_{0})\right)\nonumber\\
 &   & \quad \times 
{\partial^{2}\over (\partial M^{2})^{2}} 
\left(\frac 1 M {\partial^3\Gamma\over \partial a^3}\right)\nonumber\\
 &   &\!\!\!\!\!\!\!\! \left. +{4e^{2}\beta\over 3}
\left((\partial_{i}\hat{A}_{0})^{2}(\nabla^{2}\hat{A}_{0}) +
 (\partial_{i}\hat{A}_{0})(\partial_{j}\hat{A}_{0})(\partial_{i}\partial_{j}
\hat{A}_{0})\right){\partial^{3}\over 
 (\partial M^{2})^{3}}\left(\frac 1 M 
{\partial \Gamma\over \partial a}\right)\right]\label{c11}   
\end{eqnarray}
There are several things to note from these results. First, the
results obtained up to fifth order in derivatives
above agree completely with
the momentum space calculations. The advantage of the coordinate space
calculation lies in the fact that it directly gives us the effective
action as opposed to the momentum space calculations which give only the
amplitudes. Second, even
though the expression in Eq. (\ref{c8'}) is not manifestly gauge invariant,
terms up to fifth order in derivatives are explicitly invariant under
{\it small} gauge transformations. A gauge invariant form of the
effective action,  derived
above, needs the use of various algebraic identities and it is not
{\it a priori} clear that, at higher orders, gauge invariant
expressions will be obtained. To that end, we now give a simple proof
of {\it small} gauge invariance to all orders.

\section{Proof of invariance under {\it small} gauge transformation:}

Let us consider the effective action in Eq. (\ref{c3'}) which is linear in
$\vec{A}$. If we now make a gauge transformation,
$A\!\!\!\slash\rightarrow A\!\!\!\slash + \partial\!\!\!\slash\alpha$,
where $\alpha$ is the parameter of transformation,
then, the change in the effective action is given by
\begin{equation}
\delta\Gamma_{eff}^{(1)} = - e \sum_{n} {\rm Tr}\,{1\over
p\!\!\!\slash + 
\gamma_{0}(\tilde{\omega}_{n}+e\hat{A}_{0})+M}\,(\partial\!\!\!\slash\alpha)
\end{equation}
Let us now use the standard canonical commutation relation
$$
\left[p_{i},\alpha\right] = -i(\partial_{i}\alpha)
$$
as well as the cyclicity of trace (We note here that the zeroth order
term in this expression is the only term that needs regularization and
we have already seen that it is manifestly gauge invariant. The higher
order terms are well defined and satisfy cyclicity of trace.) to write
\begin{equation}
\delta\Gamma_{eff}^{(1)} =  ie \sum_{n} {\rm
Tr}\,\left[p\!\!\!\slash , {1\over p\!\!\!\slash + 
\gamma_{0}(\tilde{\omega}_{n}+e\hat{A}_{0})+M}\right]\,\alpha
\end{equation}
Let us next write
\begin{equation}
p\!\!\!\slash = p\!\!\!\slash +
\gamma_{0}(\tilde{\omega}_{n}+e\hat{A}_{0})+M -
(\gamma_{0}(\tilde{\omega}_{n}+e\hat{A}_{0})+M)
\end{equation}
Using this, leads to
\begin{eqnarray}
\delta\Gamma_{eff}^{(1)} & = & ie \sum_{n} {\rm
Tr}\left[-(\gamma_{0}(\tilde{\omega}_{n}+e\hat{A}_{0})+M) {1\over p\!\!\!\slash +
\gamma_{0}(\tilde{\omega}_{n}+e\hat{A}_{0})+M}\right.\nonumber\\
 &  & \;\;\left. + {1\over p\!\!\!\slash +
\gamma_{0}(\tilde{\omega}_{n}+e\hat{A}_{0})+M}(\gamma_{0}(\tilde{\omega}_{n}+e\hat{A}_{0})+M)\right]\alpha\nonumber\\ 
 & = & ie \sum_{n} {\rm
Tr}\left[-(\gamma_{0}(\tilde{\omega}_{n}+e\hat{A}_{0})+M) {1\over
p\!\!\!\slash + 
\gamma_{0}(\tilde{\omega}_{n}+e\hat{A}_{0})+M}\right.\nonumber\\
 &  & \;\;\left.(\gamma_{0}(\tilde{\omega}_{n}+e\hat{A}_{0})+M) {1\over p\!\!\!\slash +
\gamma_{0}(\tilde{\omega}_{n}+e\hat{A}_{0})+M}\right]\alpha\nonumber\\
 & = & 0
\end{eqnarray}
Here, we have used the cyclicity of the trace in the second term and
the fact  that the factor in the numerator is a
multiplicative operator which commutes with $\alpha$. This proves that
the expression that we are interested in is invariant under {\it
small} gauge transformations, even though it may not be manifest.

This, therefore, raises the question as to whether we can have a
derivative expansion which will give a  manifestly ({\it small}) gauge
invariant expression for the effective action. The answer, not
surprisingly, is in  the affirmative. Let us
recall that
\begin{equation}
\Gamma_{eff} [A,M] = -  \sum_{n} {\rm Tr}\,\ln\left(p\!\!\!\slash +
\gamma_{0}(\tilde{\omega}_{n}+e\hat{A}_{0})+M+eA\!\!\!\slash\right)
\end{equation}
In two dimensions, the vector field has the simple decomposition
$$
A_{i} = \partial_{i}\sigma + \epsilon_{ij}\partial_{j}\rho
$$
from which, it can be determined that
\begin{equation}
(\partial^{2}\rho) = -\epsilon_{ij}\partial_{i}A_{j} = - B\label{c11'}
\end{equation}
Using this decomposition and the familiar properties of gamma matrices
in two dimensions, we can write
\begin{eqnarray}
\Gamma_{eff} [A,M] & = & - \sum_{n} {\rm Tr}\,\ln\,e^{-ie\sigma}
\left(p\!\!\!\slash + e\gamma_{0}(\partial\!\!\!\slash\rho) +
\gamma_{0}(\tilde{\omega}_{n}+e\hat{A}_{0})+M\right)e^{ie\sigma}\nonumber\\
 & = & - \sum_{n} {\rm Tr}\,\ln\left(p\!\!\!\slash +
e\gamma_{0}(\partial\!\!\!\slash\rho) + 
\gamma_{0}(\tilde{\omega}_{n}+e\hat{A}_{0})+M\right)
\end{eqnarray}
From the definition of the parity violating effective action in
Eq. (\ref{c4}), it follows now that
\begin{eqnarray}
{\partial\Gamma_{eff}^{{\rm PV}}\over \partial a} & = & -{e\over 2\beta} \sum_{n}
{\rm Tr}\left[{1\over p\!\!\!\slash + e\gamma_{0}(\partial\!\!\!\slash\rho) +
\gamma_{0}(\tilde{\omega}_{n}+e\hat{A}_{0})+M}\right.\nonumber\\
 &  & \;\;\left. - {1\over p\!\!\!\slash +
e\gamma_{0}(\partial\!\!\!\slash\rho) + 
\gamma_{0}(\tilde{\omega}_{n}+e\hat{A}_{0})-M}\right]\gamma_{0}\nonumber\\
 & = & -{eM\over \beta} \sum_{n} {\rm Tr}{1\over
p^{2}+\tilde{\omega}_{n}^{2}+M^{2}+N}\gamma_{0}\label{c12}
\end{eqnarray}
where we have defined
\begin{equation}
N =
(-ie\gamma_{0}(\partial\!\!\!\slash\hat{A}_{0})+2e\tilde{\omega}_{n}\hat{A}_{0}
+ e^{2}\hat{A}_{0}^{2}) + ie\gamma_{0}(\partial^{2}\rho) -
e^{2}(\partial_{i}\rho)(\partial_{i}\rho)
\end{equation}
Using Eq. (\ref{c11'}), the last two terms in $N$ can be expressed in terms
of $B$ and while the last one has a nonlocal form in terms of $B$, the
penultimate term is local. Expression (\ref{c12}) can now  be expanded
to linear  order in the $B$  field to give
\begin{eqnarray}
{\partial\Gamma_{eff}^{{\rm PV} (1)}\over \partial a} & = &
-{e^{2}M\over \beta} 
\sum_{n}
{\rm Tr}\left[{1\over 
p^{2}+\tilde{\omega}_{n}^{2}+M^{2}+(-ie\gamma_{0}(\partial\!\!\!\slash\hat{A}_{0})+2e\tilde{\omega}_{n}\hat{A}_{0}+e^{2}\hat{A}_{0}^{2})}\right.\nonumber\\
 &  &\;\left.\times (i\gamma_{0}B) {1\over
p^{2}+\tilde{\omega}_{n}^{2}+M^{2}+(-ie\gamma_{0}(\partial\!\!\!\slash\hat{A}_{0})+2e\tilde{\omega}_{n}\hat{A}_{0}+e^{2}\hat{A}_{0}^{2})}\right]\gamma_{0}\nonumber\\
 & = &\!\! {2ie^{2}M\over \beta} \sum_{n} {\rm tr} {1\over
(p^{2}+\tilde{\omega}_{n}^{2}+M^{2}+2e\tilde{\omega}_{n}\hat{A}_{0}+e^{2}\hat{A}_{0}^{2})^{2}
- e^{2}(\partial_{i}\hat{A}_{0})^{2}}B\nonumber\\
 &  & 
\end{eqnarray}
To any order in the $\hat{A}_{0}$ fields, the denominator can be expanded in
a systematic manner as discussed earlier. However, this form has the
advantage that it is manifestly gauge invariant to begin
with. Furthermore, there are no Dirac matrices or momentum operators
in the numerator to complicate the calculation. The only complication
may be that integrating over $a$ to obtain the action may be nontrivial.

\section{General features of the effective action:}

It is clear that, while at every order the effective action can be
determined in a closed form, its structure may not be that simple. On
the other hand, from the analysis of the effective action up to fourth
order (in fields) brings out some nice features that are worth
discussing.

First, the structures in Eqs. (\ref{c7},\ref{c8},\ref{c9}) suggest
that, to  all orders (in the
$\hat{A}_{0}$ fields) the leading
order term in the parity violating part of the effective action has
the form (terms linear in the derivative)
\begin{eqnarray}
\left(\Gamma_{eff}^{PV (1)}\right)_{1} & = & i\sum_{n=0} \int
d^{2}x\,{1\over n!} B(\beta\hat{A}_{0})^{n}\,{\partial^{n}\Gamma (a,M)\over
\partial a^{n}}\nonumber\\
 & = & i\int d^{2}x\,B\,\Gamma (a+\beta\hat{A}_{0},M)\label{c13}
\end{eqnarray}
Here, the subscript refers to the number of derivatives contained in
the effective action.This gives the simple result that the leading order
correction to the static result can be obtained completely from the
static result itself. Furthermore, this action is invariant under {\it
large} gauge transformations whenever the action with $\hat{A}_{0}=0$
is. Finally, we note that, at very high temperatures,
$\beta\rightarrow 0$, so that the action reduces to (\ref{a9'}), which
is consistent with the fact that the action in (\ref{a9'}) gives the
leading terms of the parity violating action at high temperatures.

Even the next order terms in the expansion (namely, third order in
derivatives) in Eqs. (\ref{c7},\ref{c8},\ref{c10}) seem to have a nice
structure and, with a
little bit of analysis, suggest that they can be summed to a simple
form. Let us note that
\begin{eqnarray}
\left(\Gamma_{eff}^{PV (1)}\right)_{3} & = & - iM \int
d^{2}x\,B\left[{\beta\over 6} (\nabla^{2}\hat{A}_{0}) {\partial^{2}\over
\partial M^{2}\partial a}\right.\nonumber\\
 &  &\!\! + {\beta^{2}\over 12} (2(\nabla^{2}\hat{A}_{0})\hat{A}_{0} +
(\vec{\nabla}\hat{A}_{0})\cdot (\vec{\nabla}\hat{A}_{0}))
{\partial^{3}\over \partial M^{2}\partial a^{2}}\nonumber\\
 &  &\!\! + \left. {\beta^{3}\over 12}
\hat{A}_{0}((\nabla^{2}\hat{A}_{0})\hat{A}_{0} +
(\vec{\nabla}\hat{A}_{0})\cdot (\vec{\nabla}\hat{A}_{0}))
{\partial^{4}\over \partial M^{2}\partial a^{3}}+\cdots
\right]{\Gamma(a,M)\over M}\nonumber\\
 &  & \label{c14}
\end{eqnarray}
The key observation that one can make here is that once the two derivatives
have distributed over two $\hat{A}_{0}$ fields in all possible manner
(namely, $\hat{A}_{0}(\nabla^{2}\hat{A}_{0})$ and
$(\vec{\nabla}\hat{A}_{0})\cdot (\vec{\nabla}\hat{A}_{0})$),
any number of additional $\hat{A}_{0}$ fields can only occur as
multiplicative factors. Consequently, one can view them as factors of
$a$ and bring out the appropriate functional dependence. This is most
easily seen in momentum space. We can assume that only the two
$\hat{A}_{0}$ fields on which the derivatives act to have non-zero
momentum while the other multiplicative $\hat{A}_{0}$ fields to have
zero momentum. But, then, by the decomposition in
Eqs. (\ref{c1'},\ref{c1}), these can simply be thought of as factors
of $a$. Now, since the linear term in $\hat{A}_{0}$ occurs as
$(\nabla^{2}\hat{A}_{0})$ (see Eq. (\ref{c7})), while the corresponding term in
Eq. (\ref{c8}) cannot be written as $(\nabla^{2}X)$, from the
structure of the third
derivative term in Eq. (\ref{c7}), the complete effective action in
this order in the number of derivatives can easily be seen to be of the form
\begin{equation}
\left(\Gamma_{eff}^{PV (1)}\right)_{3} = i\int d^{2}x\,B\left[c_{1}
  (\nabla^{2}\hat{A}_{0}) {\partial\over \partial a} +
  c_{2}\nabla^{2}\right]{\partial\over \partial M^{2}} \left({\Gamma
  (a+\beta\hat{A}_{0},M)\over M}\right)\label{c14'}
\end{equation}
Here $c_{1},c_{2}$ are constants to be determined. The structure of
terms in Eq. (\ref{c14'}) is interesting in that we do not see any
sign of terms of the form $(\vec{\nabla}\hat{A}_{0})\cdot
(\vec{\nabla}\hat{A}_{0})$ alluded to above. This is because such a
term can always be decomposed as
\[
(\vec{\nabla}\hat{A}_{0})\cdot (\vec{\nabla}\hat{A}_{0}) =
 - (\nabla^{2}\hat{A}_{0})\hat{A}_{0} + \frac{(\nabla^{2}\hat{A}_{0}^{2})}{2}
\]
which is also the reason why we have not introduced a structure of the form
\[
i\int d^{2}x\,c_{3} B (\vec{\nabla}\hat{A}_{0})\cdot
\vec{\nabla}\,{\partial^{2}\over \partial M^{2}\partial
a}\left({\Gamma(a+\beta \hat{A}_{0},M)\over M}\right)
\]
in Eq. (\ref{c14'}), because it is not independent of the two
structures already present there. Returning to Eq. (\ref{c14'}), we
note that we can compare the
linear term in $\hat{A}_{0}$ of this expression with the corresponding
term  in Eq. (\ref{c14}), which  determines $c_{2}={c_{1}\over \beta}= -{M\over
12}$. With this, the quadratic and the cubic terms in $\hat{A}_{0}$
following from Eq. (\ref{c14'}) coincide with the corresponding terms
in Eq. (\ref{c14}). This, therefore, suggests that the complete parity
violating effective action with three derivatives can be written in
the simple form
\begin{equation}
\left(\Gamma_{eff}^{PV (1)}\right)_{3} = -{iM\over 12} \int
  d^{2}x\,B\left[\beta(\nabla^{2}\hat{A}_{0}) {\partial\over \partial a} +
  \nabla^{2}\right] {\partial\over \partial M^{2}}\left({\Gamma
  (a+\beta\hat{A}_{0},M)\over M}\right)\label{c14''}
\end{equation}
This is also
quite interesting  in that it tells us that the terms
cubic in derivatives can be determined to all orders (in the fields)
from the static action itself. Furthermore, this effective action is
invariant under {\it large} gauge transformations because of the
derivative acting on $\Gamma$.

The terms with five derivatives (see
Eqs. (\ref{c7},\ref{c8},\ref{c11})) look quite complicated, but have a
pattern to them. However, from the observation made earlier, to
determine a simple functional form, one needs all the four derivatives
to be distributed over four $\hat{A}_{0}$ fields in all possible
manner. This, however, cannot be determined at the level of the box diagram 
($B\hat{A}_{0}^{3}$ level). We need to go to higher order amplitudes
and so, it would seem difficult to determine a simple functional
relationship for the parity violating effective action with five
derivatives to all orders in the $\hat{A}_{0}$ fields from the data we
have. Nonetheless, let us note that all these terms in
Eqs. (\ref{c7},\ref{c8},\ref{c11}) are invariant under {\it large}
gauge transformations because of the derivatives acting on $\Gamma(a,M)$.

While the above formulae are suggestive, it would be nice to see if
they are indeed correct and the origin of such structures. This can be
done in the following way. Let us recall that we are interested in
evaluating the effective action (up to normalization)
\begin{eqnarray}
 \Gamma_{eff} & = & - \sum_{n} {\rm Tr}\,\ln (p\!\!\!\slash +
 eA\!\!\!\slash + \gamma_{0}(\omega_{n}+e({a\over \beta}+\hat{A}_{0}) + M)\nonumber\\
  & = & - \sum_{n} {\rm Tr}\,\ln (p\!\!\!\slash +
 eA\!\!\!\slash + \gamma_{0}\bar{\omega}_{n} + M)
\end{eqnarray}
where we have defined
\begin{equation}
\bar{\omega}_{n} = \omega_{n} + {e\over \beta}(a+\beta\hat{A}_{0}) = {(2n+1)\pi\over
\beta} + {e\over \beta}(a+\beta\hat{A}_{0})\label{c15}
\end{equation}
Following Eqs. (\ref{a4},\ref{a5}), we can now define \cite{6}
\begin{equation}
\bar{\rho}_{n} = \sqrt{\bar{\omega}_{n}^{2}+M^{2}},\qquad
\bar{\phi}_{n} = \tan^{-1}\left({\bar{\omega}_{n}\over M}\right)
\end{equation}
where $\bar{\rho}_{n}$ and $\bar{\phi}_{n}$ are now coordinate
dependent because of the presence of $\hat{A}_{0}$. The effective
action, in these variables, takes the form
\begin{eqnarray}
\Gamma_{eff} & = & - \sum_{n} {\rm Tr}\,\ln
(p\!\!\!\slash + eA\!\!\!\slash
 + \bar{\rho}_{n}e^{\gamma_{0}\bar{\phi}_{n}})\nonumber\\
 & = & - \sum_{n} {\rm Tr}\,\ln
e^{\gamma_{0}\bar{\phi}_{n}/2} (p\!\!\!\slash + eA\!\!\!\slash +
\bar{\rho}_{n} - {i\over
2}\gamma_{0}(\partial\!\!\!\slash\bar{\phi}_{n}))e^{\gamma_{0}\bar{\phi}_{n}/2}\nonumber\\
 & = & - \sum_{n} {\rm Tr}\,\ln (p\!\!\!\slash + eA\!\!\!\slash +
\bar{\rho}_{n} - {i\over
2}\gamma_{0}(\partial\!\!\!\slash\bar{\phi}_{n}))\label{det}
\end{eqnarray}
This indeed is reminiscent of the determinant factor in \cite{6}. The
Jacobian factor, however, seems to be missing. It is well understood
in the context of derivative expansion \cite{12} that the effect of
the Jacobian is hidden in the regularization used to evaluate the
determinant (another way to view this is to note that the Jacobian is
after all a determinant and product of two determinants is again a
determinant). 

The parity violating part of the effective action can be obtained from
Eq. (\ref{det}) through a derivative expansion and would have an odd number of
$\bar{\phi}_{n}$ terms. The action, as we have already shown, is gauge
invariant and, if we are interested in terms linear in $\vec{A}$,
would depend  linearly on $B$ as well as terms with derivatives
acting on $\bar{\rho}_{n}$ and $\bar{\phi}_{n}$. From the
definition of these variables, we see that $\bar{\rho}_{n}$ has the
canonical dimension of energy while $\bar{\phi}_{n}$ is
dimensionless. This allows us to organize the successive terms in the
expansion. Thus, for example, the term linear in the derivative, would
follow from this to be
\begin{eqnarray}
\left(\Gamma_{eff}^{PV (1)}\right)_{1} & = & a_{1} \sum_{n}
\int d^{2}x\,B \bar{\phi}_{n}\nonumber\\
 & = & a_{1} \int d^{2}x\, B \tan^{-1}\left(\tanh{\beta
M\over 2} \tan ({e(a+\beta\hat{A}_{0})\over 2})\right)
\end{eqnarray}
Here, $a_{1}$ is a constant which can be fixed by requiring that in
the limit $\hat{A}_{0}=0$, we would like the effective action to
reduce to the static action. This determines $a_{1} =  {ie\over 2\pi}$
so that, we obtain
\begin{equation}
\left(\Gamma_{eff}^{PV (1)}\right)_{1} = i\int
d^{2}x\,B\,\Gamma(a+\beta\hat{A}_{0},M) 
\end{equation}
This is, of course, the action we had derived earlier for the leading
corrections to the static action in Eq. (\ref{c13}).

At the order of terms cubic in the derivatives, let us note that
the most general term we can write for the parity violating effective
action will have the form
\begin{eqnarray}\label{gen1}
\left(\Gamma_{eff}^{PV (1)}\right)_{3} & = &  \sum_{n}
\int d^{2}x\,B\left[b_{1} {(\nabla^{2}\bar{\phi}_{n})\over
\bar{\rho}_{n}^{2}} + b_{2} {(\vec{\nabla}\bar{\rho}_{n})\cdot
(\vec{\nabla}\bar{\phi}_{n})\over \bar{\rho}_{n}^{3}}\right]\nonumber\\
 & = & \sum_{n} \int
d^{2}x\,B\left[eMb_{1}\left({(\nabla^{2}\hat{A}_{0})\over
\bar{\rho}_{n}^{4}} - {2e\bar{\omega}_{n}\over \bar{\rho}_{n}^{6}}
(\vec{\nabla}\hat{A}_{0})\cdot
(\vec{\nabla}\hat{A}_{0})\right)\right.\nonumber\\
 &   &\qquad \qquad \left. + e^{2}Mb_{2}\, {\bar{\omega}_{n}\over
\bar{\rho}_{n}^{6}}\, (\vec{\nabla}\hat{A}_{0})\cdot
(\vec{\nabla}\hat{A}_{0})\right]
\end{eqnarray}
It is clear that the contribution of the first term starts with terms
of the type $B\hat{A}_{0}$ while the second structure has contribution
starting with $B\hat{A}_{0}^{3}$. Consequently, the coefficients
$b_{1}$ and $b_{2}$ can be identified from our earlier calculations
and take the values $b_{1}= {ie\over 12\pi}, b_{2}=0$. Thus, the
parity violating part of the effective action which is cubic in the
derivatives can be written as
\begin{eqnarray}
\left(\Gamma_{eff}^{PV (1)}\right)_{3} & = & {ie^{2}M\over 12\pi}
\sum_{n} \int d^{2}x\,B\left[{(\nabla^{2}\hat{A}_{0})\over
(\bar{\omega}_{n}^{2}+M^{2})^{2}} - {2e\bar{\omega}_{n}\over
(\bar{\omega}_{n}^{2}+M^{2})^{3}} (\vec{\nabla}\hat{A}_{0})\cdot
(\vec{\nabla}\hat{A}_{0})\right]\nonumber\\
 & = & -{ie^{2}M\over 24\pi} \sum_{n} \int d^{2}x\,B
{\partial\over \partial M^{2}}\left[{(\nabla^{2}\hat{A}_{0})\over
(\bar{\omega}_{n}^{2}+M^{2})} + {1\over
eM}(\nabla^{2}\tan^{-1}{\bar{\omega}_{n}\over M})\right]\nonumber\\
 &\!\!\!\!\!\!\!\! = & \!\!\!\!\!\!\!\!-{iM\over 12} \int
d^{2}x\,B\left[\beta (\nabla^{2}\hat{A}_{0}){\partial\over \partial a} +
\nabla^{2}\right] {\partial\over \partial M^{2}} \left({\Gamma
(a+\beta\hat{A}_{0}, M)\over M}\right)
\end{eqnarray}
Once again, this coincides with the expression we had determined
earlier in Eq. (\ref{c14''}).

If we go to the next order, namely terms containing five derivatives,
there are twelve possible structures that arise. However, let us note
from the terms with three derivatives that terms with
$\bar{\phi}_{n}$ occur with at least $\nabla^{2}$ acting on them.
[This corresponds to finding $b_{2}=0$ in Eq. (\ref{gen1})].
If we guess that a similar pattern continues to hold at orders higher
than the box amplitude (this could be checked by an explicit
calculation of the six point amplitude), the
parity violating effective action with five derivatives is uniquely
determined by the behaviour of the two and four point amplitudes, and
has the simple form
\begin{eqnarray}
\left(\Gamma_{eff}^{PV (1)}\right)_{5} & = & -{ie\over 60\pi}
\sum_{n} \int d^{2}x\,B\left[{\nabla^{4}\bar{\phi}_{n}\over
\bar{\rho}_{n}^{4}} -
{3(\nabla^{2}\bar{\phi}_{n})(\nabla^{2}\bar{\rho}_{n})+4(\partial_{i}\bar{\rho}_{n})
(\partial_{i}\nabla^{2}\bar{\phi}_{n})\over
\bar{\rho}_{n}^{5}}\right.\nonumber\\
 &  & \qquad\qquad \left. +
{7(\nabla^{2}\bar{\phi}_{n})(\partial_{i}\bar{\rho}_{n})^{2}\over
\bar{\rho}_{n}^{6}}\right]
\end{eqnarray}
Upon doing the sum over the discrete frequencies, this determines 
the following form for the corresponding all orders (in fields) effective
action
\be
\begin{array}{lll}
\left(\Gamma_{eff}^{PV (1)}\right)_{5} & = & 
-\displaystyle{
{iM\over 30} \int {\rmd}^{2}x\,B
\left\{e^{2}\beta^{2}
\left[
(\partial_i\hat A_0)^2\right]^2\left(\frac{13}{8}+\frac{5}{12}M^2
\frac{\partial}{\partial M^2}\right)
\frac{\partial}{\partial a}\frac{\partial}{\partial M^2}
\right.} 
\\ & & \\ & \!\!\!\!\!\!\!\!\!\!\!\!\!\!\!\!\!\!\!\! \!\!\!\!\!\!\!\!\!\!+& 
\!\!\!\!\!\!\!\!\!\!\!\!\!\!\!\!\!\!\!\!\!\!\!\!\!\!\!\!\!\!
e^{2}\beta\left[\displaystyle{(\partial_i\hat A_0)^2 \nabla^2\hat A_0
\left(\frac{11}{2} + \frac 5 3 M^2 \frac{\partial}{\partial
    M^2}\right)}\right.
\\ & & \\ & \!\!\!\!\!\!\!\!\!\!\!\!\!\!\!\!\! \!\!\!\!+& 
\!\!\!\!\!\!\!\!\!\!\!\!\!\!\!\!\!\!\!\!\!
\left.\displaystyle{
\partial_i\hat A_0 \partial_j\hat A_0 \partial_i\partial_j \hat A_0
\left(\frac{20}{3}+2M^2 \frac{\partial}{\partial M^2}\right)}\right]
\displaystyle{\frac{\partial}{\partial M^2}}
\\ & & \\ & \!\!\!\!\!\!\!\!\!\!\!\!\!\!\!\!\!\!\!\!\!\!\!\!\!\!\!\!\!\!-& 
\!\!\!\!\!\!\!\!\!\!\!\!\!\!\!\!\!\!\!\!\!\!\!\!\!\!\!\!\!\!
\displaystyle{\left.
\beta^{2}\left[\partial_i\hat A_0\partial_i\nabla^2\hat A_0+
\frac{5}{12}(\nabla^2\hat A_0)^2+
\frac 1 3(\partial_i\partial_j\hat A_0)^2\right]
\frac{\partial}{\partial a} - \frac{\beta}{2}
(\nabla^{2})^{2}\hat{A}_{0}\right\}}
\\ & & \\ &
\!\!\!\!\!\!\!\!\!\!\!\!\!\!\!\!\!\!\!\!\!\!\!\!\!\!\!\!\!\!\times & 
\!\!\!\!\!\!\!\!\!\!\!\!\!\!\!\!\!\!\!\!\!\!\!\!\!\!\!\!\!\!
\displaystyle{ \qquad 
\frac{\partial}{\partial a} \left(\frac{\partial}{\partial M^2}\right)^2
\left(\frac{\Gamma(a+\beta\hat A_0,M)}{M}\right)}
\end{array}.
\ee
In the above expression, all the derivatives act only on the field
immediately to their right.

This discussion makes it clear that such an analysis can be carried
out systematically to any order in the derivatives (of course, one
needs to calculate higher point functions), which, in turn, would
determine the corresponding all order effective action. Interestingly,
all such effective actions can be determined completely from a
knowledge of the leading order parity violating action in the static
limit.

\section{Conclusion:}   

In this paper, we have tried to go beyond the leading order term in
the static parity violating effective action in the case of the
Abelian Chern-Simons theory using the derivative expansion. We have
discussed the various subtleties that arise in using derivative
expansion in such a theory and have improved and extended the earlier
proposed  method \cite{13} for calculating the leading order term in
this approach. We have shown, in this approach, that the parity
violating effective action, in the static limit, is linear in the
$\vec{A}$ field. In going beyond the leading order, we have used the
momentum space method to calculate the self-energy and the box
diagrams up to fifth order in the momenta. We have used the derivative
expansion in the coordinate space to determine the parity violating
effective action up to fourth order in fields. All these actions can
be obtained in closed form (namely, powers of derivatives can be
summed) in principle. However, their forms are neither very illuminating
nor useful. In contrast, at any given order of the derivatives, we can
sum the effective action containing all possible fields. The resulting
effective actions are determined completely by the leading order
action in the static limit. We have given a proof of {\it small} gauge
invariance of the effective action. We have also shown that all the
higher order terms are {\it large} gauge invariant. We have tried to
discuss the possible origin of the interesting structure of the higher
order terms that arise in the derivative expansion.

This work was supported in part by US DOE Grant No. DE-FG 02-91ER40685
and by CNPq and Fapesp, Brazil.

\providecommand{\href}[2]{#2}\begingroup\raggedright

\endgroup

\end{document}